\documentclass[aps,superscriptaddress,nofootinbib,eqsecnum,pra,notitlepage,twocolumn]{revtex4-1} 

\usepackage{amssymb}
\usepackage{amsmath,amsfonts,amsthm,bm}
\usepackage{graphicx,xcolor}
\usepackage{float}
\usepackage{hyperref}
\usepackage{subfigure}
\usepackage{dcolumn}
\usepackage{soul}
\usepackage{ulem}
\usepackage{verbatim}

\DeclareMathAlphabet{\mathcal}{OT1}{pzc}{m}{it}
\DeclareMathOperator{\csch}{csch}

\begin{document}

\title{Single-particle momentum distribution of Efimov states in noninteger dimensions}

\author{D. S. Rosa}

\affiliation{Instituto Tecnol\'{o}gico de Aeron\'{a}utica, DCTA,
  12228-900 S\~{a}o Jos\'{e} dos Campos, SP, Brazil}
  
  \author{T. Frederico}

\affiliation{Instituto Tecnol\'{o}gico de Aeron\'{a}utica, DCTA,
  12228-900 S\~{a}o Jos\'{e} dos Campos, SP, Brazil}

\author{G. Krein}

\affiliation{Instituto de F\'isica Te\'orica, Universidade Estadual Paulista,
Rua Dr. Bento Teobaldo Ferraz, 271-Bloco II, 01140-070 S\~ao Paulo, SP, Brazil}

\author{M. T. Yamashita}

\affiliation{Instituto de F\'isica Te\'orica, Universidade Estadual Paulista,
Rua Dr. Bento Teobaldo Ferraz, 271-Bloco II, 01140-070 S\~ao Paulo, SP, Brazil}

\begin{abstract}
We studied the single-particle momentum distribution of mass-imbalanced
Efimov states embedded in noninteger dimensions. The contact parameters, 
which can be related to the thermodynamic properties of the gas, were 
calculated from the high momentum tail of the single particle densities. 
We studied the dependence of the contact parameters with the progressive 
change of the noninteger dimension, ranging from three (D=3) to two (D=2) 
dimensions. Within this interval, we move from the (D=3) regime where the 
Efimov discrete scale symmetry drives the physics, until close to the critical 
dimension, which depends on the mass imbalance, where the continuum scale 
symmetry takes place. We found that the two- and three-body contacts grow
significantly in magnitude with the decrease of the noninteger dimension 
towards the critical dimension, impacting observables of resonantly interacting 
trapped Bose gases.
\end{abstract}

\maketitle

\section{Introduction}

More than 20 years of advances in cold-atom technologies allowed not only the 
experimental confirmation of Efimov states in homo~\cite{homoexp0,homoexp1,homoexp2} 
and heteronuclear atomic systems~\cite{heteexp1,heteexp2,heteexp3} but also  
lead to the burst of the rich research area of Efimov Physics. 
 Nowadays, atomic traps give remarkable freedom to conveniently manipulate energies 
and geometries to study several aspects of few-body physics. Magnetic 
fields tuned to a Feshbach resonance~\cite{feshbach} allow controlling the value of 
scattering lengths, and asymmetrical magnetic fields 
allow squeezing atomic clouds to create tri-~\cite{BEC3D}, two-~\cite{BEC2D}, and one-dimensional~\cite{BEC1D} 
environments.

The achievement of the universal regime,  in that the scattering 
length tends to infinity, effectively  made possible the experimental
confirmation of Efimov states{\textemdash}weakly-bound systems originally predicted 
by Vitaly Efimov in 1970 when studying three identical bosons~\cite{efimov0,efimov1}. 
This effect is characterized by the three-boson system exhibiting 
an infinite number of geometrically spaced energy levels (see Refs.~\cite{
Braaten:2004rn,Naidon:2016dpf,Greene:2017cik,
Hammer:2019poc} for reviews), and was first observed through indirect 
measurement of the three-body loss peaks in trapped cold atomic 
systems~\cite{homoexp3}. Nowadays, advanced  experimental techniques 
allow the direct measurement of the binding energies of two- and three-body 
molecules in cold atomic gases~\cite{fletcher0,musolino}.

A myriad of developments followed since the experimental confirmation 
of Efimov states. One of the most notable developments is the measurements 
(e.g. those of Refs.~\cite{fermions1,makotyan,fletcherscience,Yan2020,Zou2021}) 
of the Tan's contacts~\cite{Tan1,Tan2,Tan3}, remarkable universal quantities 
that parameterize thermodynamic relations between macroscopic observables 
such as the momentum distribution, energy, and response functions of 
low-temperature gases interacting via short-range interactions. Tan found 
those universal relations by studying the tail of the two-body momentum distribution 
of unitary Fermi gases. For unitary Bose gases, there is one more contact parameter~\cite{castindensity}, related to the probability of finding three atoms 
close together.  

Since a long time, it is known the determinant 
role played by the spatial dimension
for the presence of the Efimov effect in a three-body 
system{\textemdash}it is present in three dimensions but
absent in two~\cite{lim1,lim2}. As a consequence, 
in two dimensions  physical properties of few- and many-boson 
systems  scale with the 
two-body energies in the limit of zero-range interactions~\cite{adhikari2d}.
Tan's contacts, in particular also reflect such a
scaling~\cite{bellotti2d}. Therefore, the sequential disappearance 
of the most excited bound states during a progressive change in the 
effective dimension of a confined resonant three-body system should 
also have consequences for Tan's contacts.

An efficient way to 
study a dimensional crossover is to introduce a continuous dimension~$D$ and 
solve the three-body problem employing only the inter-atomic interactions, 
with the $D$-dependent centrifugal barrier mocking the external squeezing potential~\cite{D3,D4,BOD,PRARD,garridoconection1}. 
Although technically convenient to implement, connecting $D$ to an 
experimental setup is a key issue.  For three identical bosons in 
a deformed trap induced by an external harmonic potential, such a
connection was suggested in Ref.~\cite{garridoconection3} to be: 
\begin{equation}
\frac{3(D-2)}{(3-D)(D-1)} = \left(\frac{b_{ho}}{r_{2D}}\right)^2 ,
\label{eq:Dtrap}
\end{equation} 
where $b_{ho}$ is the oscillator length and $r_{2D}$ is the root-mean-square 
radius of the bound three-body system in two dimensions. A similar
expression for a two-body system was suggested in Ref.~\cite{garridoconection1}.

Despite the advances that led to the possibility of compressing and 
expanding atomic clouds, creating effectively two-~\cite{BEC2D} and 
one-dimensional~\cite{BEC1D} setups, to the best of our knowledge, 
there  are not yet experiments designed to 
study the effects of continuous deformation of the
trap on Efimov physics. While awaiting such an experimental possibility, 
the rich physics revealed by previous theoretical studies warrants 
exploring this subject in connection with Tan's contacts. Within 
such a perspective, we study in this work the $D$-dependence of 
Tan's two- and three-body contact parameters in mass-imbalanced three-body 
systems featuring the Efimov effect.  We extract the contact parameters 
from the single-particle momentum distributions at high momentum values. 
We treat the three-body problem in terms of $D$-dimensional hyperspherical coordinates~\cite{reportD} 
and solve the problem analytically using the Bethe-Peierls (BP) boundary conditions employing 
the method, we introduced in Ref.~\cite{betpeiPRA}. 

This work is organized as follows. In section~\ref{section2}, for a system composed by two atoms $A$ and a third one $B$, we review the derivation of the analytical $D$-dimensional Faddeev components of the mass-imbalanced three-body bound state wave function. Section~\ref{section3} is devoted to the derivation of the momentum distribution of particle $B$ in $D$-dimensions. We also discuss in this section, the high momentum regime of the single particle momentum distribution from where the two- and three-body contacts are obtained. Section~\ref{sec:quantitative} shows quantitative results for the momentum density and relate them with the two- and three-body contacts for three-identical bosons and a mass-imbalanced system of the form $^6$Li$-^{133}$Cs$_2$. The conclusions are given in Section~\ref{section4}. Appendices~\ref{appn1} to \ref{appn4} give details of the large momentum sub-leading contributions to the single particle momentum distribution discussed in Sec.~\ref{section3}.

 \section{$D$-dimensional Efimov state }
 \label{section2}

In this section, we review the derivation of the $D$-dimensional three-body wave function 
of an Efimov state for a mass-imbalanced system at unitarity, according to our
approach introduced in Ref.~\cite{betpeiPRA}. We found the solution of the energy 
eigenvalue equation for a three-particle system interacting with a zero-range potential by considering the Bethe-Peierls boundary condition~\cite{bethe} on the 
free energy eigenstate. This method uses the fact that the short-range 
region, where the interaction strongly affects the wave function, can be neglected as only 
the asymptotic region, parametrized by the scattering  length, is the relevant one.

%
\subsection{Configuration space}

We consider three different particles with masses $m_i$, $m_j$, $m_k$, and coordinates $\textbf{x}_{i}$, $\textbf{x}_{j}$ and $\textbf{x}_{k}$. One can eliminate the center of mass coordinate and describe the system in terms of two relative Jacobi coordinates. 
The three sets of such coordinates are given by
\begin{equation}
 \mbox{\boldmath$r$}_{i} = \textbf{x}_{j} - \textbf{x}_{k}\quad\text{and}\quad
 \mbox{\boldmath$\rho$}_{i} = \textbf{x}_i - \frac{m_j\textbf{x}_j+m_{k}\textbf{x}_k}{m_j + m_k} \, ,
\end{equation}
where ($i, j, k$) are taken cyclically among ($1,2,3$). The Faddeev decomposition of the three-body wave function 
allows to write it as a sum of three components. In the center of mass, it reads: $$\Psi(\textbf{x}_{1},\textbf{x}_{2},\textbf{x}_{3}) = 
\psi^{(1)}(\mbox{\boldmath$r$}_1,\mbox{\boldmath$\rho$}_1) + 
\psi^{(2)}(\mbox{\boldmath$r$}_2,\mbox{\boldmath$\rho$}_2)
+ \psi^{(3)}(\mbox{\boldmath$r$}_3,\mbox{\boldmath$\rho$}_3)\,.$$ 
Each Faddeev component satisfies the free
Schr\"{o}dinger eigenvalue equation: 
\begin{equation}\label{eq:schr3B}
\left[\frac{1}{2\eta_{i}}\nabla^{2}_{\mbox{\boldmath$r$}_i} +\frac{1}{2\mu_{i}} 
\nabla^{2}_{\mbox{\boldmath$\rho$}_i}
- E_3\right] \psi^{(i)} (\mbox{\boldmath$r$}_i,\mbox{\boldmath$\rho$}_i) = 0,
\end{equation}
where  $E_3$ is the energy eigenvalue. The reduced masses are given by
 $\eta_{i} = m_{j}m_{k}/(m_{j}+m_{k})$ and $ \mu_{i} = {m_{i}(m_{j}+m_{k})}/({m_{i}+m_{j}+m_{k}})$.
For convenience, we simplify the form of the kinetic energies introducing the new 
coordinates $
 \mbox{\boldmath$r$}'_{i} = \sqrt{\eta_i}\, \mbox{\boldmath$r$}_{i}\quad$ and $\quad
 \mbox{\boldmath$\rho$}'_{i} = \sqrt{\mu_i}\, \mbox{\boldmath$\rho$}_{i}$. The three sets of 
 primed coordinates are related to each other by the orthogonal transformations
\begin{eqnarray}
\mbox{\boldmath$r$}'_{j}& =& - \mbox{\boldmath$r$}'_{k}\cos\theta_i + \mbox{\boldmath$\rho$}'_{k}
\sin \theta_i, \nonumber \\
\mbox{\boldmath$\rho$}'_{j}& =& - \mbox{\boldmath$r$}'_{k}\sin\theta_i - 
\mbox{\boldmath$\rho$}'_{k}\cos \theta_i,
\end{eqnarray}
where $\tan \theta_i = \left[m_i M/(m_j\ m_k)\right]^{1/2}$, with $M = m_1 + m_2 + m_3$. 

Considering three distinct bosons in a state with vanishing total angular momentum, 
one can write a reduced Faddeev component as $
\chi^{(i)}_0 (r'_{i}, \rho'_i) = \left( r'_{i} \ 
\rho'_{i}\right)^{ (D-1)/2} \psi^{(i)}(r'_i,\rho'_i)$.  
The solution of the corresponding eigenvalue equation for $\chi^{(i)}_0$ is found 
by using  hyperspherical coordinates to separate the variables 
$r'_i = R \sin \alpha_i$ and $\rho'_i = R \cos \alpha_i $, so that one can write $
\chi^{(i)}_0(R,\alpha_{i})  = C^{(i)} F(R)\,G^{(i)}(\alpha_{i})$,
where $R^{2}= r_i^{\prime 2}+\rho_i^{\prime2}$ and $\alpha_i = \arctan(r'_i/\rho'_i)$. The functions
$F(R)$ and $G^{(i)}(\alpha_{i})$ satisfy the following differential equations:
\begin{eqnarray}
&&\left[- \frac{\partial^{2}}{\partial R^{2}} + \frac{s_{n}^{2}-1/4}{R^{2}} + 2\kappa_0^2
\right]\sqrt{R}F(R)=0,
 \label{radialwavefunc}
\\
&&\left[- \frac{\partial^{2}}{\partial \alpha_i^{2}} -s_{n}^{2}+\frac{(D-1)(D-3)}{ \sin^2 2 
\alpha_i}\right] G^{(i)}(\alpha_i)=0, 
\label{Eq:angularD}
\end{eqnarray}
where $-\kappa_0^2 = E_3$ and $s_n$ is recognized as the Efimov parameter.

The definitions $z = \cos 2\alpha_i$ and $G^{(i)} = (1-z^2)^{1/4} g^{(i)}$ turn
Eq.~\eqref{Eq:angularD} into the associated Legendre differential 
equation~\cite{legendrebook} with known analytical solutions:
\begin{eqnarray} 
G^{(i)}(\alpha_i) &=& \sqrt{\sin2 \alpha_i}\Big[ P_{s_n/2-1/2}^{D/2-1}\,(\cos2\alpha_i) \nonumber \\
&-& \frac{2}{\pi}\tan\big[\pi(s_{n} -1)/2\big] Q_{s_n/2-1/2}^{D/2-1}\,(\cos2\alpha_i)\Big],\ \ \ \ \
\label{Eq:AngSol}
\end{eqnarray}
where $P_{n}^{m}(x)$ and $Q_{n}^{m}(x)$ are the associated Legendre functions. A finite value for the Faddeev component $\psi^{(i)}$ at $\rho_i =0$ imposes that $G^{(i)}(\alpha_i=\pi/2)= 0$, since $\rho_i' = R \cos{\alpha_i}$.

\begin{widetext}

Considering the solution of the hyperradial
equation~\eqref{radialwavefunc}, and the hyperangular eigenfunction, 
Eq.~\eqref{Eq:AngSol}, each Faddeev component of the wave function  is
written as~\cite{betpeiPRA}:
\begin{eqnarray}
\psi^{(i)}(r'_i,\rho'_i) &=&C^{(i)}   \frac{ K_{ s_n}\left(\sqrt{2} \kappa_0 \sqrt{  r'^{2}_{i}+  \rho'^{2}_{i} } 
\right) }
{ \big(  r'^{2}_{i}+ \rho'^{2}_{i} \big)^{D/2-1/2}}\frac{\sqrt{\sin\big[2 \arctan\left( 
r'_i/\rho'_i\right)\big]}}{\big\{\cos\big[ \arctan\left( r'_i/\rho'_i\right)\big]\ \sin\big[ \arctan\left( 
r'_i/\rho'_i\right)\big]\big\}^{D/2-1/2}}
\nonumber \\
&\times&\left[ P_{s_n/2-1/2}^{D/2-1}\Big\{\cos\big[2 \arctan( 
r'_i/\rho'_i)\big]\Big\}-\frac{2}{\pi}\tan\big[\pi(s_n-1)/2 \big] Q_{s_n/2-1/2}^{D/2-1}\Big\{\cos\big[2 
\arctan( r'_i/\rho'_i)\big]\Big\}\right]\, ,
\label{wavefunction}
\end{eqnarray}
where $K_{ s_n}$ is the modified Bessel function of the second kind.

The Efimov parameter $s_n$ is obtained considering that all three pairs of 
particles interact resonantly. The BP boundary condition at the unitary
limit~\cite{betpeiPRA} must be satisfied by the three-body wave 
function when each relative distance between two of the particles tends to zero. 
Taking the three cyclic permutations of $\{i,j,k\}$ one has a homogeneous linear system: 
\begin{equation} 
\frac{C^{(i)}}{2}
\left[ \left(\cot\alpha_i\right)^{\frac{D-1}{2}} 
\left( \sin2\alpha_i \frac{\partial}{\partial \alpha_i} 
+ D-3\right) G^{(i)} (\alpha_i) \right]_{\alpha_i\rightarrow 0}
+ (D -2) \left[ \frac{C^{(j)} \, G^{(j)}(\theta_k)}
{\left(\sin\theta_k \cos\theta_k\right)^{\frac{D-1}{2}}} 
+ \frac{C^{(k)} \, G^{(k)}(\theta_j)}
{\left(\sin\theta_j \cos\theta_j\right)^{\frac{D-1}{2}}} 
\right] = 0\,,
\label{BPsystem}
\end{equation} 
for $i\neq j \neq k$. 
\end{widetext}
The Efimov parameter, $s_n$, is obtained by solving the characteristic transcendental equation of the system. When $s_n$ is purely imaginary ($s_n \rightarrow is_0$), the effective $1/R^{2}$ potential in  Eq.~(\ref{radialwavefunc}) is attractive, giving rise to the well known Landau ``fall-to-center", where the energy spectrum is unbounded from below{\textemdash}a  behavior first found by by Thomas~\cite{thomas} for a 
neutron-deuteron nuclear system.

\subsection{Momentum space} \label{subsect:momentumspace}

To obtain the single particle momentum distributions,
one needs to perform the Fourier Transform (FT) of the Faddeev wave functions, Eq.~\eqref{wavefunction}. However, instead of performing  the FT directly, 
we  first obtain the spectator amplitude. The asymptotic form of 
the associated Legendre polynomials for $r'_i\to 0$ in the hyperangular
part of the Faddeev wave function, Eq.~\eqref{Eq:AngSol}, has to be used, 
which leads to:
\begin{eqnarray}
\psi^{(i)}\left(\rho'_i,r'_i \right)&\underset{r'_i\to 0}{=}&C^{(i)} \frac{\sqrt{2} \left[1-i\cot \left(D \pi/ 2\right) \tanh \left(s_0 \pi/2  \right)\right]}{\Gamma \left(2-D/2\right)} \nonumber \\
&\times &\ {r'_i}^{2-D}\frac{K_{i s_0}\left(\sqrt{2}\kappa_0 \rho'_{i} \right)}{\rho'_i}, 
\label{wfrlimit}
 \end{eqnarray}
where $\Gamma(z)$ is the gamma function defined for all complex numbers $z$, 
except for the non-positive integers{\textemdash}this condition restricts the 
validity of our results to the interval $2\leq D <4$. 

The spectator function, namely $B^{(i)}(\rho_i)$, can be found by taking advantage 
of the fact that each Faddeev component $\psi^{(i)}(\rho'_i,r'_i)$ obeys the Schr\"odinger’s equation for the contact interaction, written as:
\begin{equation}
\left[\nabla_{r'_i}^{2}+\nabla_{\rho'_i}^{2}- 2\kappa_0^2 \right] \psi^{(i)}(r'_i,\rho'_i)= \delta(r'_i)B^{(i)}(\rho'_i)\,.
\label{freeschroe}
\end{equation}
We substitute Eq.~\eqref{wfrlimit} in~\eqref{freeschroe}, which, in the limit  $r'_{i}\to0$, gives:
\begin{eqnarray}
B^{(i)}(\rho'_i)&=&C^{(i)}\frac{2^{3/2} \pi^{D/2}\left[1-i\cot \left(D\pi/ 2\right) \tanh \left(s_0 \pi/2  \right)\right]}{\Gamma(D/2)\Gamma \left(2-D/2\right)} \nonumber \\
&\times&\frac{K_{i s_0}\left(\sqrt{2}\kappa_0 \rho'_{i} \right)}{\rho'_i}.
 \label{specfuncrho}
\end{eqnarray}
Now, we take the $D$-dimensional FT
\begin{equation}
\int d^{D}\rho'_i \exp(-i \textbf{q}'_i.\bm{\rho}'_i)B^{(i)}(\rho'_i) = \chi^{(i)}(q'_i)\,,
\end{equation}
to write the spectator function in momentum space ($q'_i=  q_i/\sqrt{\mu_i}$) as: 
 \begin{multline}
 \chi^{(i)}(q'_i)=C^{(i)}\mathfrak{F}_{(D,s_0)}
\kappa_{0}^{1-D} \\ \times
H_2  \tilde{F}_1 \left(\mathcal{F}_{(D,s_0)}^{*} ,\mathcal{F}_{(D,s_0)},\frac{D}{2},-\frac{q_i^{\prime 2}}{2\kappa_0^2}
 \right),
 \label{regulatespect}
\end{multline} 
where
\begin{eqnarray}
\mathfrak{F}_{(D,s_0)}& \equiv & i \frac{2^{D/2+1} \pi^{D-1}\  \Gamma \big[ \mathcal{F}_{(D,s_0)}\big]\,
 \Gamma \big[ \mathcal{F}_{(D,s_0)}^{*}\big] }{(D-2)}\nonumber \\
 &\times&\cos \left(\frac{\pi}{2}(D-i s_0)  \right)\csch \left( \frac{\pi}{2} s_0\right)
\end{eqnarray}
and $H_2 \tilde{F}_1(a,b,c,z)$ is the regularized hypergeometrical function with $\mathcal{F}_{(D,s_0)} \equiv (D-1+is_0)/2$. 

\begin{center}
\begin{figure}[t] 
{\includegraphics[width=8.2cm]{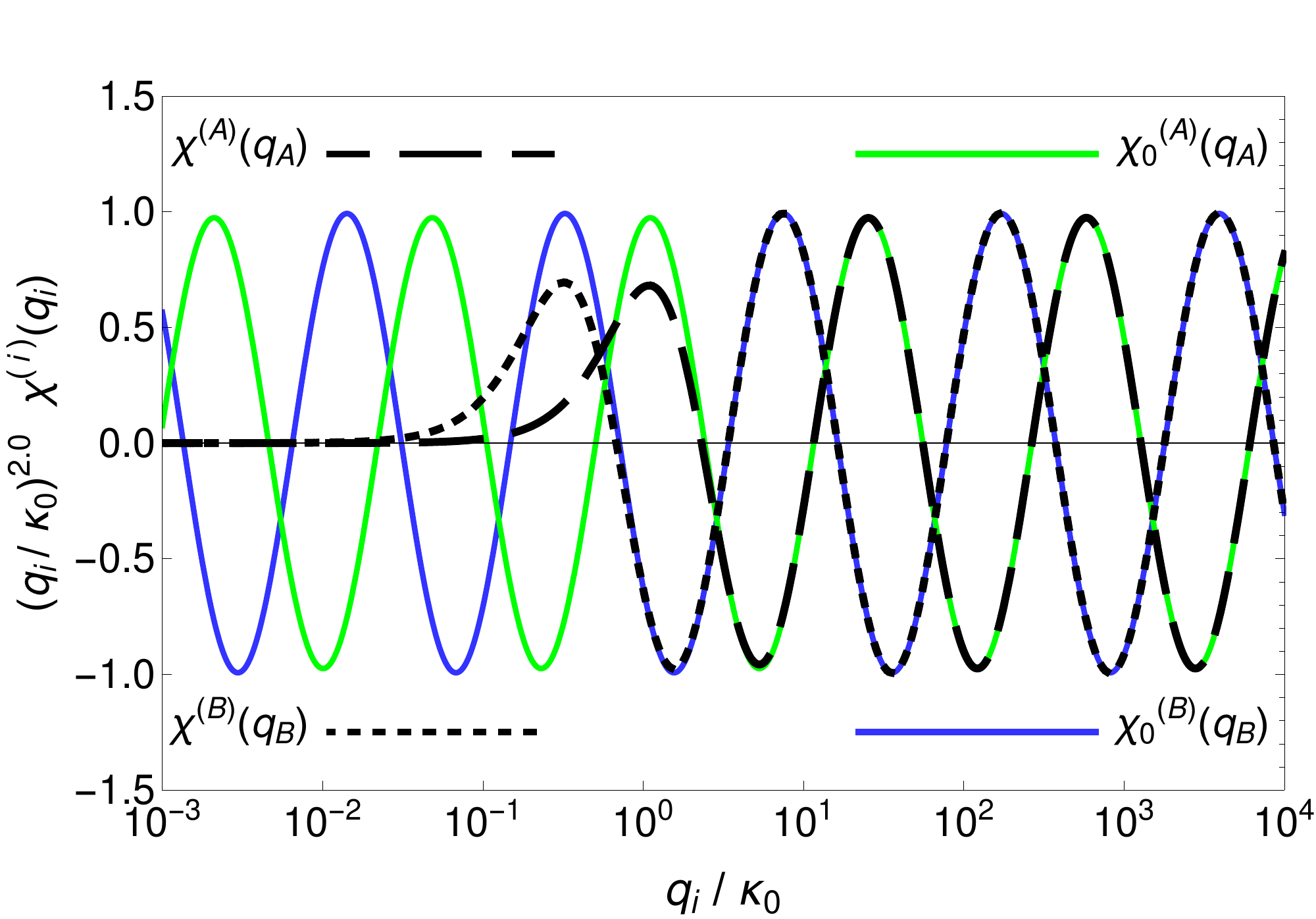}} 
{\includegraphics[width=8.2cm]{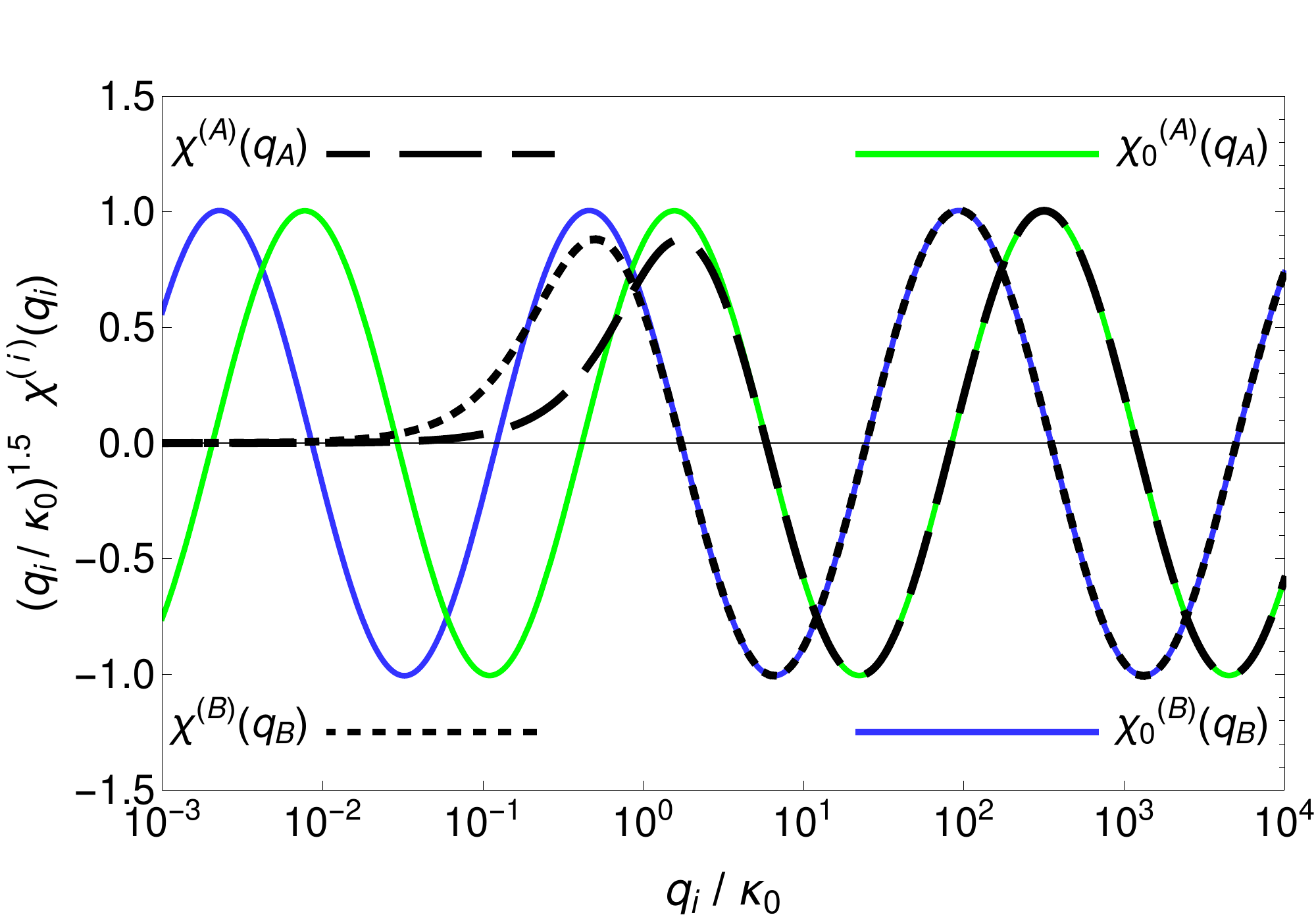}}
\caption{{ Spectator functions in momentum space for the $^{6}$Li$-^{133}$Cs$_{2}$ system with  finite three-body energy, $\chi^{(i)}(q_i)$  
($i=A\equiv\,^{133}$Cs or $B\equiv^{6}$Li), computed with Eq.~\eqref{regulatespect} for
$\chi^{(A)}(q_A)$ (long-dashed line) and $\chi^{(B)}(q_B)$ (short-dashed line), compared to the zero-energy case  from Eq.~\eqref{eq:assymp} for $\chi^{(A)}_0(q_A)$ (green solid line) and $\chi^{(B)}_0(q_B)$ (blue solid line). Top: three dimensions.  
Bottom: $D=2.5$, which corresponds to a harmonic-trap length of $b_{ho}/r_{2D}=\sqrt{2}$.} 
}
\label{fig1}
\end{figure}
\end{center}

The characteristic log-periodic behavior of the spectator functions exhibited 
in the Efimov region is present in the asymptotic form of the spectator functions, which is found from Eq.~\eqref{regulatespect} at large momentum as 
\begin{eqnarray}
 \chi^{(i)}_0(q'_i) &=&C^{(i)}\mathfrak{F}_{(D,s_0)}
 2\sqrt{\operatorname{Re}(\mathcal{G})^2+\operatorname{Im}(\mathcal{G})^2} \nonumber \\
& \times&\left(\frac{q'_i}{\sqrt{2}} \right)^{1-D}\cos\left[ s_0 \ln \left(\frac{q'_i}{\sqrt{2}\kappa_0^{*}} \right)\right]\,,\ \ \ 
 \label{eq:assymp}
\end{eqnarray}
where $\kappa_0^{*}\equiv \kappa_0/\exp\left\{\arctan\left[{\operatorname{Im}(\mathcal{G})/\operatorname{Re}(\mathcal{G})}\right]/s_0\right\}$ and 
\begin{equation}
 \mathcal{G}  
=\frac{\Gamma\left(\mathcal{F}_{(D,s_0)}-\mathcal{F}_{(D,s_0)}^{*} \right)}{\Gamma\left(\mathcal{F}_{(D,s_0)}-D/2-1\right)\Gamma\left(\mathcal{F}_{(D,s_0)} \right)}\,.
\label{eq:G}
\end{equation}
We note that the asymptotic form of the spectator function in Eq.~\eqref{eq:assymp} also corresponds to the limit of vanishing the three-body energy, 
with $\kappa^*_0$  associated with the necessary three-body scale parameter, which is chosen to match Eq.~\eqref{eq:assymp} with the large momentum behavior of Eq.~\eqref{regulatespect} for the finite energy spectator function. The normalization constants are solutions of Eq.~\eqref{BPsystem}, namely, the linear homogeneous system that determines the Efimov parameter.

Figure~\ref{fig1} shows the spectator functions, Eq.~\eqref{regulatespect}, 
compared to the zero energy case, Eq.~\ref{eq:assymp} conveniently normalized to one for an $AAB$ system with $A=\,^{133}$Cs and $B=^{6}$Li embedded in two different dimensions, namely $D=3$ (top  panel) 
and $2.5$ (bottom panel). 
In the low momentum region, the damping of the spectator amplitude with respect to the zero-energy case is an effect of the finite three-body binding energy. The impact of changing the dimension in which the $^{6}$Li$-^{133}$Cs$_{2}$ system 
is embedded is manifested mainly in the different log-periodicity 
of the spectator functions. The period increases to infinity as 
the system approaches the critical dimension, $D = 2.231$, for which the Efimov state disappears. The increasing separation of the log-periodic nodes towards the critical dimension is illustrated by comparing the top and bottom panels of Fig.~\ref{fig1}, when the $^{6}$Li$-^{133}$Cs$_{2}$ system  is forced to decrease from three to 2.5 dimensions, respectively. We reproduce analytically the numerical results obtained in Ref.~\cite{yamashita2013} for $D=3$.

\section{Momentum distribution }
 \label{section3}

In this section, we compute the momentum distribution of the particle $B$ for $AAB$ 
systems at the unitary limit in $D$-dimensions.  We recall that particle $B$ is 
the one responsible for  giving rise to an effective 
Efimov-like potential in the limit of heavy $A$'s, as we have  shown 
in Ref.~\cite{BOD}. The momentum densities for particle $B$ were analytically calculated for 
$D=3$ in Ref.~\cite{yamashita2013}.

We start by defining $\textbf{k}_\alpha$ ($\alpha=i,j,k$) as the momenta of each particle
in the rest frame. We have that the Jacobi momenta from one particle to the center of mass of the other two and the relative momentum of the pairs are given, respectively, by
\begin{small}
\begin{equation}
\hspace{-.15cm}\textbf{q}_i =\mu_i \left(\frac{\textbf{k}_i}{m_i}-\frac{\textbf{k}_k+\textbf{k}_k}{m_j+m_k} \right)\,\,\text{and}\,\,\textbf{p}_i = \eta_i\left(\frac{\textbf{k}_j}{m_j} - \frac{\textbf{k}_k}{m_k} \right).
\end{equation}
\end{small} 

In the following, we define the single-particle momentum distribution for particles of types 
$A $ and $B$. The Faddeev components of the three-body wave function for a zero-range interacting system, composed 
of two identical particles $A$ and a third one $B$, can be written using the FT of Eq.~\eqref{freeschroe} and the spectator function given by Eq.~\eqref{regulatespect}.

We start writing the $AAB$ bound state 
wave function in the basis $| \textbf{q}_B \textbf{p}_B \rangle $:
\begin{eqnarray}
\label{densityB}
&&\hspace{-0.75cm} 
\langle \textbf{q}_B \textbf{p}_B | \Psi \rangle = \frac{1}{E_3+p_B^2/2\eta_{B} + q_B^2/2\mu_{B}} 
\Bigl[ \chi^{(B)} (\textbf{q}_B)  \nonumber \\ 
&& + \chi^{(A)}\left(\bigl\vert\textbf{p}_B - \frac{\textbf{q}_B}{2}\bigr\vert\right)
+ \chi^{(A)}\left(\bigl\vert\textbf{p}_B + \frac{\textbf{q}_B}{2}\bigr\vert\right) \Bigr], 
\end{eqnarray}
and in the basis $| \textbf{q}_A \textbf{p}_A \rangle $:
\begin{eqnarray} 
\label{densityA}
&&\hspace{-0.75cm} 
\langle \textbf{q}_A \textbf{p}_A | \Psi \rangle = \frac{1}{E_3+p_A^2/2\eta_{A} +q_A^2/2\mu_{A}} 
\left[ \chi^{(A)} (\textbf{q}_A) \right. \nonumber \\ 
&+&\left. \chi^{(B)}\left(\bigl\vert\textbf{p}_A \!-\! \frac{\textbf{q}_A}{1+\mathcal{A}^ {-1}}\bigr\vert\right)
+ \chi^{(A)} \left(\bigl\vert\textbf{p}_A \!+\! \frac{\textbf{q}_A}{1+\mathcal{A}}\!\mid\!\right)\! \right].\nonumber \\
\end{eqnarray} 
Here, we  are using $m_A=1$ in the mass ratio $\mathcal{A} = m_B/m_A$. 

The momentum distributions in $D$-dimensions for particles $A$ and $B$ are given, respectively, by:
\begin{equation}
 n_A(q_A) = \int d^{D}p_A \ |\langle \textbf{q}_A \textbf{p}_A | \Psi \rangle |^{2} \,,
 \end{equation}
 and
\begin{equation}
\label{totdenB}
 n_B(q_B) = \int d^{D}p_B \ |\langle \textbf{q}_B \textbf{p}_B | \Psi \rangle |^{2} \,.
 \end{equation}
 
The $AAB$ wave function can be determined,  up to an 
overall constant,  by obtaining the coefficients of the spectator functions from the 
solution of the homogeneous linear system~\eqref{BPsystem}. We use the following 
normalization condition: 
\begin{equation} 
 \int d^Dq_B\, n_B(q_B)
 =1\,\,\text{or}\, \,
 \int d^Dq_A\, n_A(q_A)
 =1
 \, .
 \label{norm}
 \end{equation}
 
From Eqs.~(\ref{densityB}) and (\ref{totdenB}), we can split the momentum density 
into nine terms, which can be reduced to four, considering the symmetry between 
the two identical particles $A$. This simplifies the computation of the momentum
density to four contributions:
 \begin{equation}
 \label{4sum}
 n_B(q_B) = n_1(q_B) + n_2(q_B) + n_3(q_B) + n_4(q_B),
 \end{equation}
each of which is given by

 \begin{eqnarray}
&&n_{1}(q_B)= \lvert \chi^{(B)}(q_B) \rvert^{2} \int d^{D}p_B \frac{1}{\left(E_3 + 
 p_{B}^{2}+q_{B}^{2} \frac{\mathcal{A}+2}{4\mathcal{A}}\right)^{2}}, \ \ \ \ \ \nonumber \\
 \label{n1}
 \end{eqnarray}
 \begin{eqnarray}
 n_{2}(q_B) = 2 \int d^{D}p_B \frac{\lvert \chi^{(A)}(\lvert \textbf{p}_B -\textbf{q}_B/2 \rvert) \rvert^{2}}
 { \left( E_3 + p_{B}^{2}+ q_{B}^{2}\frac{\mathcal{A}+2}{4\mathcal{A}} \right)^{2} }, \hspace{2.1cm}
 \label{n2}
\end{eqnarray}
 \begin{eqnarray}
 n_{3}(q_B) &=& 2  \chi^{(B)}\overset{*}{(}q_B) \int d^{D}p_B \frac{\chi^{(A)}(\lvert \textbf{p}_B - \textbf{q}_B/2  \rvert )  }
 { \left( E_3 + p_{B}^{2} + q_{B}^{2}  \frac{\mathcal{A}+2}{4\mathcal{A}} \right)^{2} },\ \ \ \ \ \ \nonumber \\
 &+& {\rm c.c.},
 \label{n3}
 \end{eqnarray}
 \begin{eqnarray}
 n_{4}(q_B) &=& \int d^{D}p_B \frac{\chi^{(A)}\overset{*}{(}\lvert \textbf{p}_B-\textbf{q}_B/2 \rvert)
 \chi^{(A)}(\lvert \textbf{p}_B + \textbf{q}_B/2  \rvert )  }
 {\left( E_3 + p_{B}^{2} + q_{B}^{2}  \frac{\mathcal{A}+2}{4\mathcal{A}} \right)^{2}} \nonumber \\
 &+& {\rm c.c.}.
 \label{n4}
 \end{eqnarray}
 
Our task now is to evaluate the integral expressions in Eqs.~\eqref{n1}-\eqref{n4}  and
extract the contacts from the large-momentum tail of the distribution densities. 
The contribution $n_1(q_B)$ is straightforward to calculate: 
 \begin{eqnarray}
 n_{1}(q_B) &=&  \frac{\lvert \chi^{(B)}(q_B) \rvert^{2}}{q_{B}^{4-D}} \mathcal{S}_{D}    \frac{\pi}{4}\csc\left( \frac{D\pi}{2} \right) (2-D)\nonumber \\
 &\times&\left(\frac{\mathcal{A}+2}{4\mathcal{A}}\right)^{D/2-2} \, ,
 \end{eqnarray} 
where $\mathcal{S}_{D}$ is the area of a $D$-dimensional 
sphere. The second contribution, $n_2(q_B)$, can be computed 
from Eq.~\eqref{n2} making the change of variables $\textbf{p}_B
-\textbf{q}_B/2=\textbf{q}_A$ as:
\begin{equation}
 n_{2}(q_B) = 2 \int d^{D}q_A \frac{\lvert \chi^{(A)}(q_A) \rvert^{2}}
 { \left(  q_A^{2}+ \textbf{q}_A.\textbf{q}_B + q_{B}^{2}  \frac{\mathcal{A}+1}{2\mathcal{A}} \right)^{2} }\, .
 \end{equation}
In order to identify the leading order term in the large momentum region, we perform the 
manipulation:
\begin{eqnarray}\label{eq:n2}
 n_{2}(q_B) &=& 2 \int d^{D}q_A\lvert \chi^{(A)}(q_A) \rvert^{2}\nonumber \\
 &\times&\left[ \frac{1}
 { \left( q_A^{2}+ \textbf{q}_A.\textbf{q}_B + q_{B}^{2}  \frac{\mathcal{A}+1}{2\mathcal{A}} \right)^{2} }-\frac{4 \mathcal{A}^2}{(\mathcal{A}+1)^2}\frac{1}{q_B^4}\right] \nonumber \\
 & +& \frac{C_2}{q_B^{4}}\, ,
 \end{eqnarray}
 where $C_2$ is the two-body contact, given by
 \begin{eqnarray}
 C_{2}
= \frac{8 \mathcal{A}^2}{(\mathcal{A}+1)^2}\mathcal{S}_D \int^\infty_0 dq_A \, q_A^{D-1}\lvert \chi^{(A)}(q_A) \rvert^{2}\,.  \label{eq:c2}
 \end{eqnarray}
The contact $C_2$ can be related to the derivative w.r.t. the scattering length of the 
 gas's mean energy (or  mean free energy
at nonzero temperature). It has dimension  
$(\text{length})^{D-4}$ and therefore  scales as $C_2\propto \kappa_0^{4-D}$.
 
From the integral representations in  Eqs.~\eqref{n1}-\eqref{n4}, we obtain the oscillatory and non-oscillatory contributions to each of the four components of the momentum density at large momentum (detailed calculations of the sub-leading contributions for $n_1$ to $n_4$ at large momentum can be found in Appendices~\ref{appn1} to \ref{appn4}). The leading and sub-leading contributions in the asymptotic region are given by:
\begin{eqnarray}
&&n_B(q_B)= \frac{C_{2}}{q_{B}^{4}}  + \frac{C^{'}_{3}}{q_{B}^{D+2}} 
\nonumber \\
&& +  \frac{C_{3}}{q_{B}^{D+2}} \cos\!\left[2 s_0 \log\left(\frac{q_B/\kappa_0^{*}}
{(4\mu_A \mu_B )^{1/4} }  \right) \! + \!\Phi \right] + \cdots, \ \ \ 
\label{c3c3l}
\end{eqnarray} 
where $C_3^{'}$ brings the known non-oscillatory behavior alongside with $C_2$, $C_3$ and $\Phi$, which are, respectively, the amplitudes and the phase related to the log-periodic oscillatory term. The parameter $C_3$ is the three-body 
contact, closely related to the Efimov effect as it gives the amplitude 
of the log-periodic function of the momentum distribution. 

The contact parameter $C_3$ and the phase $\Phi$ of the log-periodic asymptotic density, Eq.~\eqref{c3c3l}, are computed by adding Eqs.~\eqref{eq:A3}, \eqref{eq:B7},  \eqref{eq:C4} and \eqref{eq:D4}. $C'_3$ is obtained by adding  Eqs.~\eqref{eq:A4}, \eqref{eq:B8}, \eqref{eq:C5} and \eqref{eq:D5}. Both parameters $C_3$ and $C'_3$ scale with $\kappa^2_0$ or, equivalently, the three-body bound-state energy.

\section{Quantitative Results} \label{sec:quantitative}

In this  section, we present  the numerical results  for the momentum density  in noninteger dimensions computed from Eq.~\eqref{4sum} with the exact spectator function~\eqref{regulatespect}, as well as the sub-leading  contributions  to  the density given by Eq.~\eqref{c3c3l}. We provide examples for real systems, and the contact parameters $C_2$, from Eq.~\eqref{eq:c2}, $C_3$  and $C'_3$ are compared with known results for three dimensions. 

\subsection{Momentum density} 

\begin{center}
\begin{figure}[t] 
{\includegraphics[width=8cm]{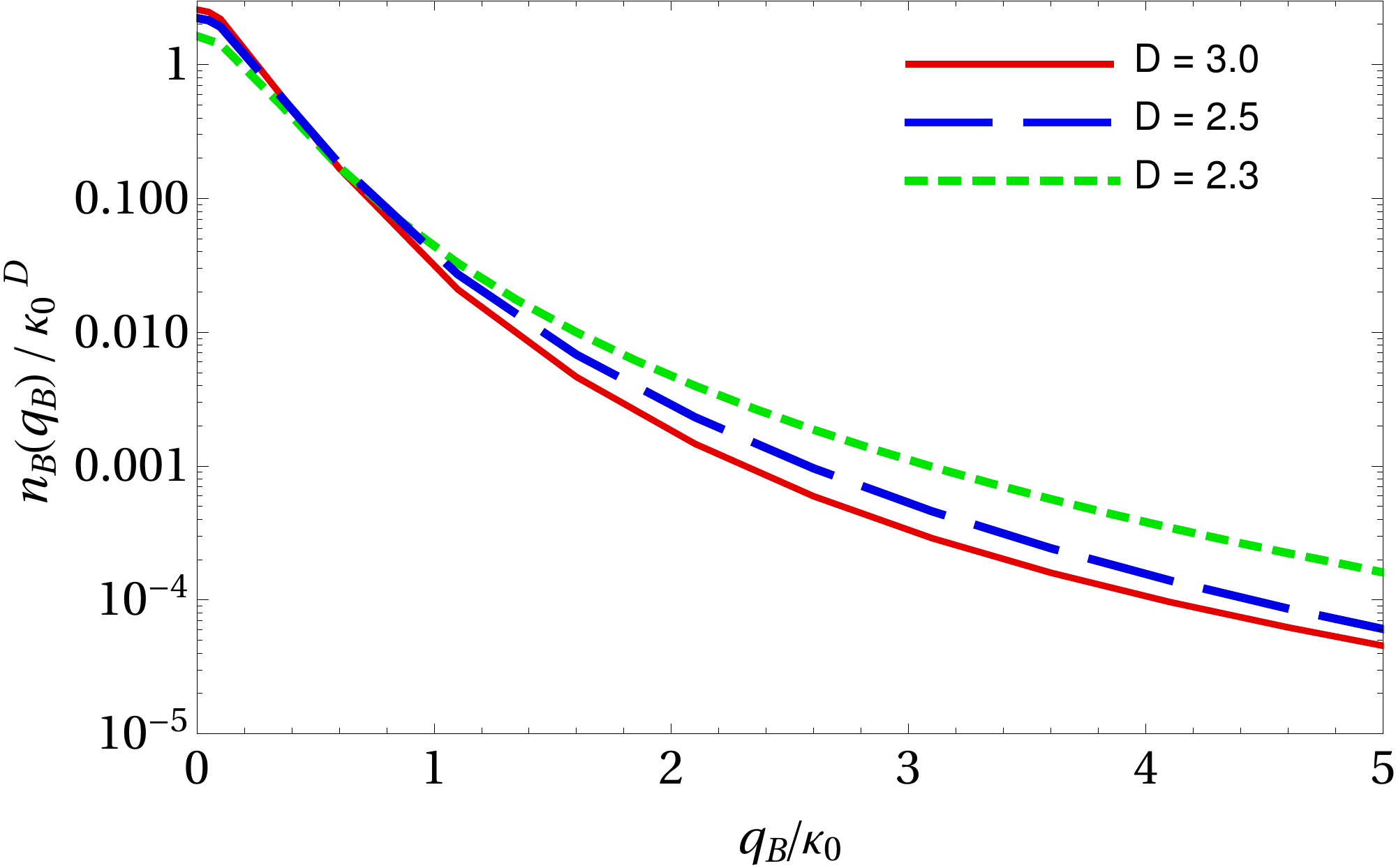}}
\caption{Single particle momentum distribution, $ n_B(q_B)$ of an $^{6}$Li$-^{133}$Cs$_{2}$  Efimov state in $D=3$ (solid line), $D=2.5$ (long-dashed line) and $D=2.3$ (short-dashed line). }
\label{fig2}
\end{figure}
\end{center}

 \begin{center}
\begin{figure}[t] 
{\includegraphics[width=8cm]{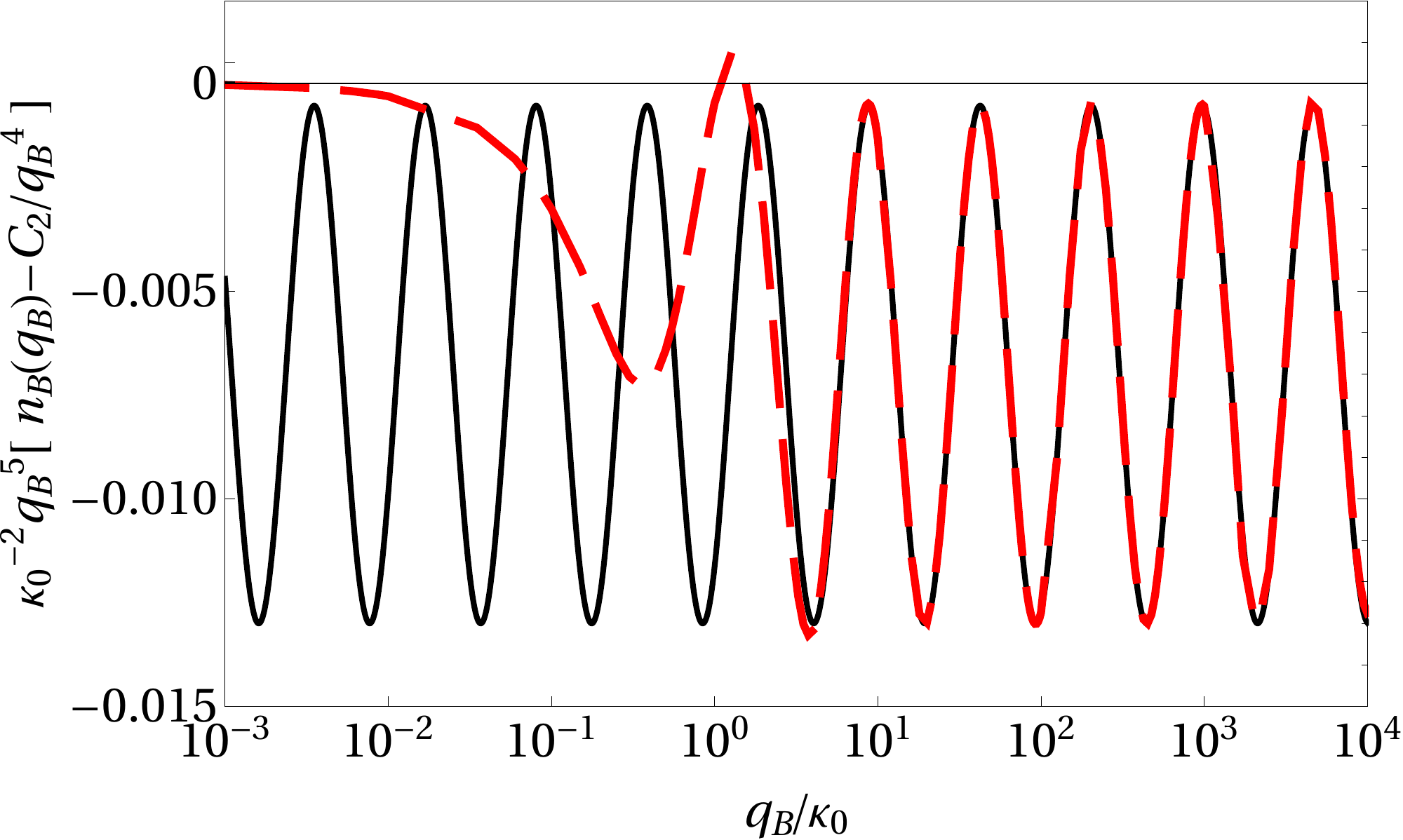}} 
{\includegraphics[width=8cm]{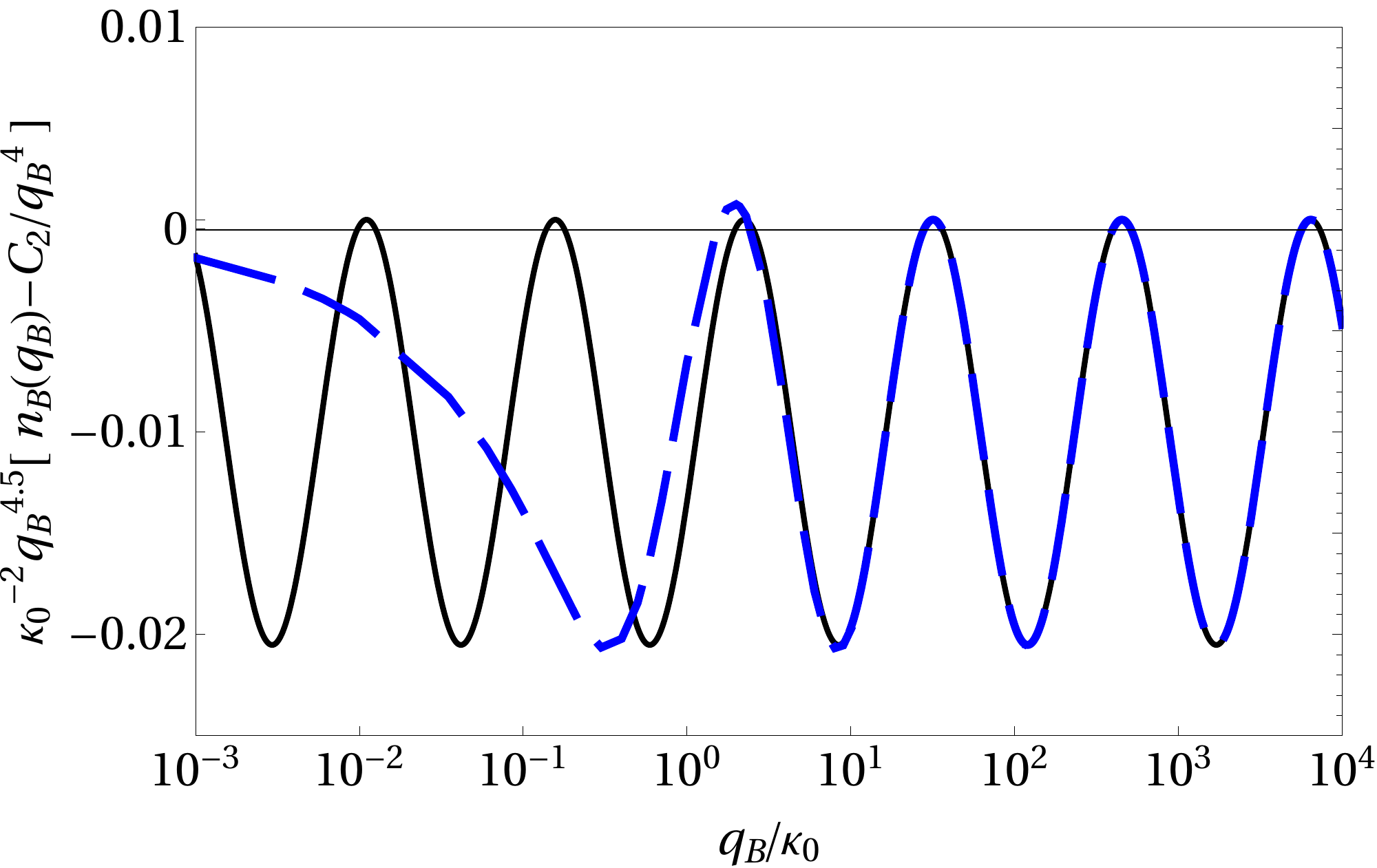}}
{\includegraphics[width=8cm]{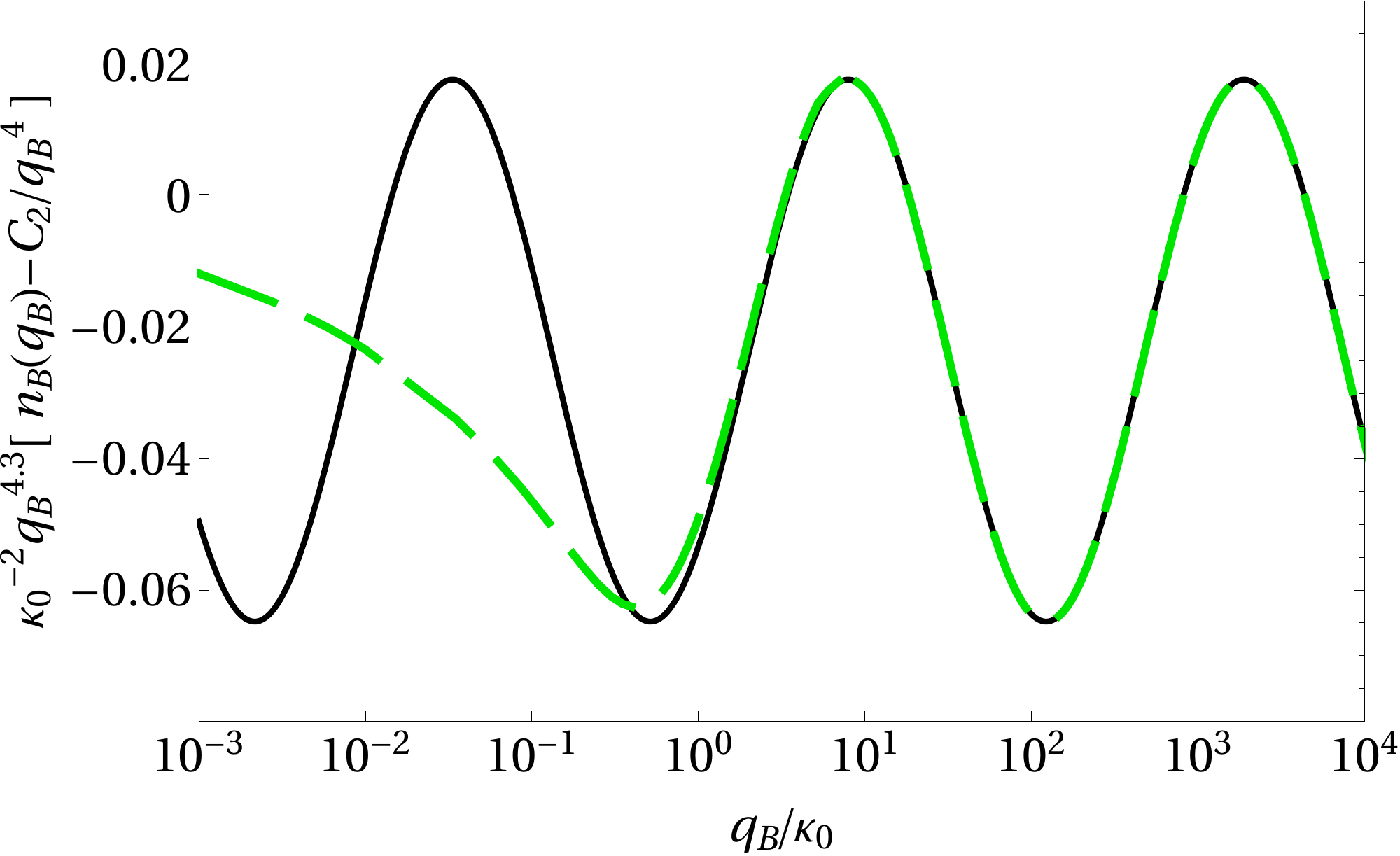}}
\caption{Subtracted single-particle momentum distribution, $ n_B(q_B)-
 {C_{2}}/{q_{B}^{4}}$, of an $^{6}$Li$-^{133}$Cs$_{2}$  Efimov state in $D=3$ (top panel), $D=2.5$ (middle panel) and $D=2.3$ 
 (bottom panel).  Results obtained with regular spectator function Eq.~\eqref{regulatespect} (dashed lines)  and subtracted asymptotic formula obtained with Eq.~\eqref{eq:assymp} (solid lines). }
\label{fig3}
\end{figure}
\end{center}

The normalized momentum density, $n_B(q_B)$, is shown for the low momentum region in Fig.~\ref{fig2}, considering the $^{6}$Li$-^{133}$Cs$_{2}$ system in 3, 2.5 and 2.3 dimensions. The results  for $D=2.3$ situate close to the critical dimension where it takes place the transition between the regimes of the Efimov discrete scale symmetry to the  continuum one.  We observe that the squeezing of the system, by lowering the noninteger dimension, tends to emphasize the large momentum region or short-distances - this, reflects naively to a well-known result in two dimensions: any weak-attractive potential is enough to bind the system for the lowest angular momentum state.

Consequently, the large momentum region is privileged, which is also expressed by the enhancement of the momentum density and the associated two and three-body contacts. This becomes evident in the figure when one follows the decrease of the noninteger dimension by observing that density is depleted close to $q_B=0$ and enhanced for larger values of $q_B/\kappa_0$. What is visible in the figure is essentially the tail $C_2/q_B^4$, which indicates that $C_2$ increases considerably from three to the critical dimension, where the Efimov effect vanishes.

In Fig.~\ref{fig3}, we show the results for the  subtracted single-particle momentum distribution ($ n_B(q_B)- C_{2}/q_{B}^{4}$)
for an Efimov state of  the $^{6}$Li$-^{133}$Cs$_{2}$  system in 3 (top panel), 2.5 (middle panel) and 2.3 (bottom panel) dimensions. The results are obtained from computing Eq.~\eqref{4sum} with the exact spectator function~\eqref{regulatespect}. These results  are compared with the sub-leading terms in  the asymptotic expansion given in Eq.~\eqref{c3c3l}, and we found that the asymptotic region is  reached quite fast and the condition $q_B \gg\kappa_0$ can be relaxed to  $q_B \gtrsim\kappa_0$. 

Comparing the top, middle and bottom panels, we observe the increasing separation between the nodes of the momentum distribution for 3,  2.5 and 2.3 dimensions, tending to infinity as the system approaches the critical dimension, where the Efimov effect disappears. The wavelength associated with the log-periodicity at  large momentum is directly related to the value of the Efimov parameter for each  noninteger dimension, such that it diverges towards the critical dimension where $s_0\to0$. 

It is also possible to observe in Fig.~\ref{fig3} that the amplitude of the log-periodic oscillations raises by decreasing the dimension from three to 2.3. Such effect corresponds to the enhancement of $C_3$  by lowering the noninteger dimension, mimicked by  strengthening the three-dimensional confinement of the system in one direction. Furthermore, we notice that the mean value reflected in $|C'_3|$ also increases. In what follows, we will   further  explore the dependence of the contact parameters, by changing the noninteger dimension, in two systems: $^{6}$Li$-^{133}$Cs$_{2}$ and three-identical bosons. 

\begin{table}[b]
\centering
\caption{ Comparison with the contacts  in three-dimensions obtained in Refs.\cite{castindensity,yamashita2013}. Note that the results from Ref.~\cite{castindensity} were multiplied by a factor of $3(2\pi)^{3}$ to agree with the normalization from Eq.~\eqref{norm}. } 
\begin{tabular}{|c|c|c|c|c|}
\hline\hline
& & & & \\
$m_B/m_A$ & Contacts   &  ~~Ref.~\cite{yamashita2013}~~  &  ~~Ref.~\cite{castindensity}~~  & our work \\ & & & & \\
\hline 
& & & & \\
 & $C_2$  &   0.0274  & - & 0.0301
\\ & & & & 
\\ 6/133 &
$C_3$      &  -    & - & 0.0062 
\\ & & & & \\
& $C_3'$ & -   & - &  -0.0067   \\ & & & & \\
& $\Phi $ & -   & - & -4.5201
\\ & & & & 
\\ \hline & & & &
\\
& ~~$C_2$~~        & 0.0715   & 0.0713 & 0.0713 \\
& & & & \\ 1 &
$C_3$      &  -    & 0.1199 & 0.1199
\\ & & & & \\
& $C_3'$ & 0   & 0 & 0 \\ & & & & \\
& $\Phi $ & -   & -0.8728 & -0.8728
\\ & & & & \\
\hline \hline
\end{tabular} 
\label{tab:1}
\end{table}

\subsection{Contact parameters}

\begin{center}
\begin{figure}[t] 
{\includegraphics[width=8.cm]{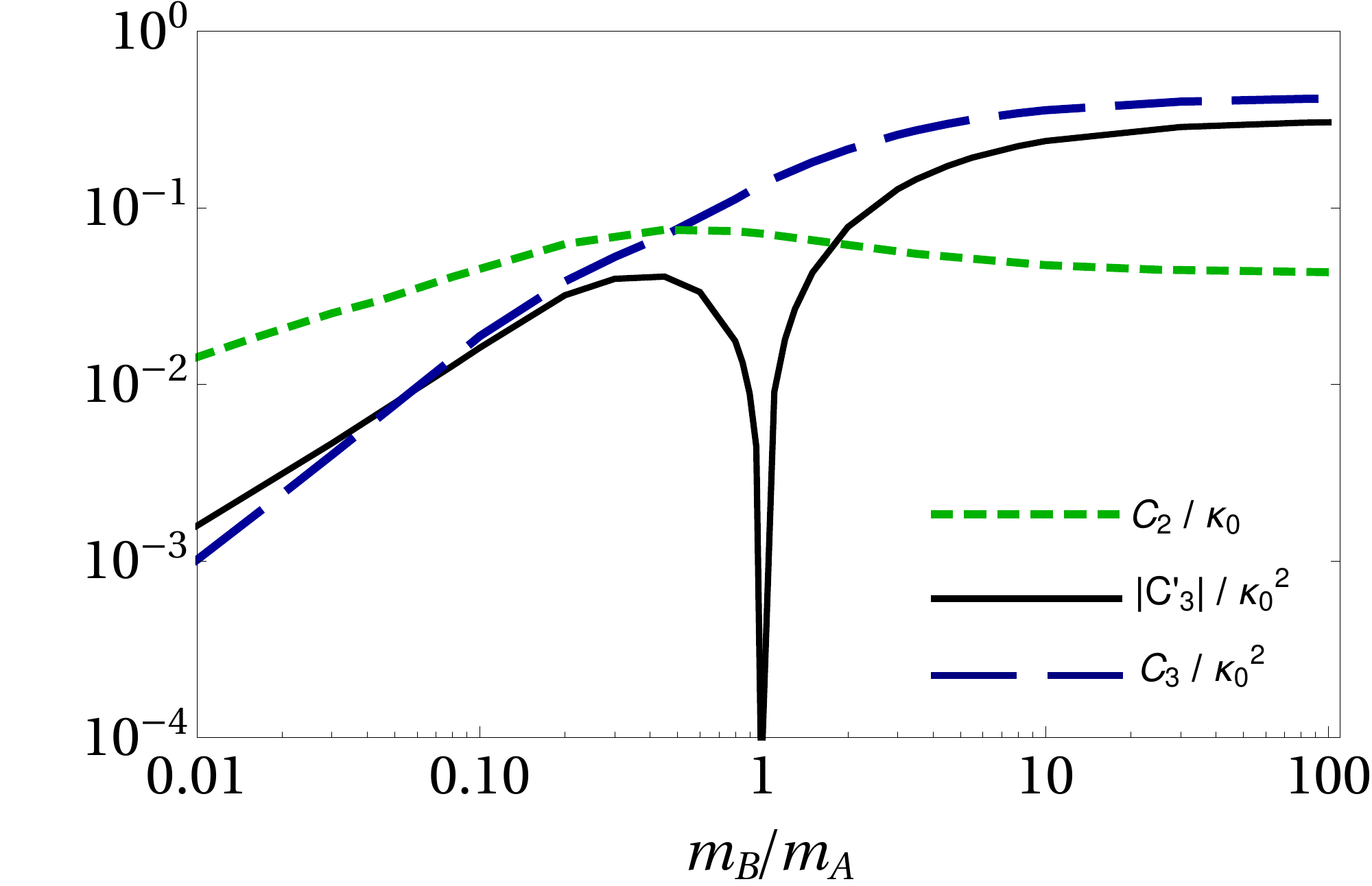}} 
{\includegraphics[width=8.cm]{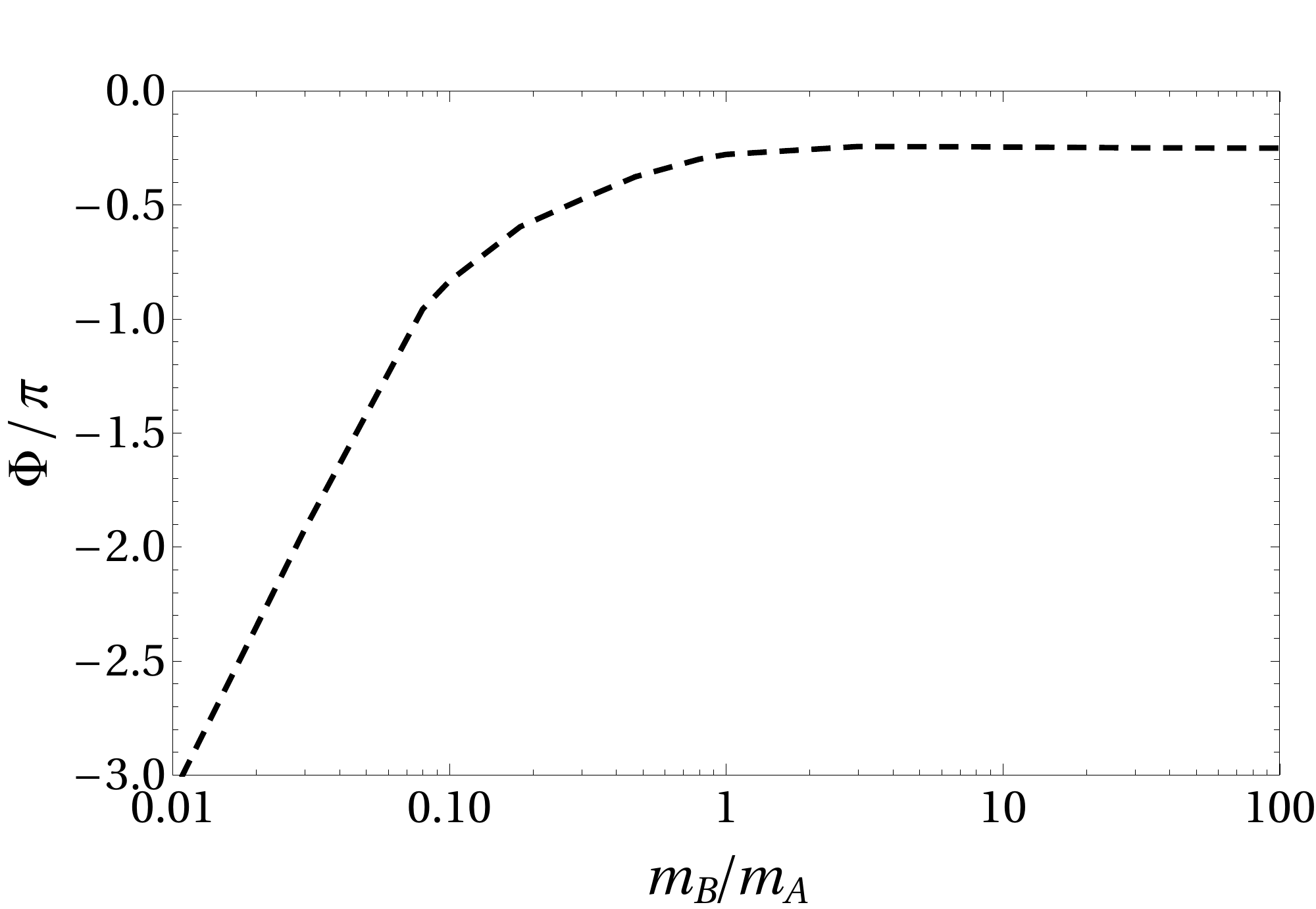}}
\caption{Three- and two-body contact parameters (top panel) and phase (bottom panel), considering a $AAB$   system with different mass ratios embedded in three dimensions.}
\label{fig4}
\end{figure}
\end{center}

We start by presenting our results for the contacts and phase for three-particle systems in three dimensions. In Table~\ref{tab:1}, we compare our calculations with results from Ref.~\cite{yamashita2013} for $m_B/m_A=6/133$ and from Ref.~\cite{castindensity} for $m_B/m_A=1$. In Fig.~\ref{fig4},  we illustrate  the dependence of $C_2$, $C_3$, $C'_3$ and $\Phi$ for different mass ratios, ranging from heavy-heavy-light to light-light-heavy systems. In the top panel of the figure, we show results for the two and three-body parameters, $C_2$, $C_3$ and $C_3^{'}$, considering a wide range of mass imbalanced  three-body systems. In the bottom panel of Fig.~\ref{fig4}, we show the results for the phase ($\Phi$) 
of the sub-leading log-periodic term in the asymptotic form of the momentum density. We observe that $C_3$, $C'_3$ and $\Phi$ saturate for the light-light-heavy system $(m_B/m_A\gg 1)$, as well as $C_2$.  Furthermore, for $(m_B/m_A\gg 1)$, the 
heavy particle $B$ tends to be closer to the center of mass of the $AAB$ system, 
increasing the probability to find it in the large momentum region, which is 
reflected in the large values of the contacts, while on the other extreme for $(m_B/m_A\ll 1)$, where the Efimov parameter increases and the density for the light-particle $B$ becomes very diffuse turning less probable to find it at small distances, and consequently the contacts decrease.
 
We should observe that the three-body parameter $C_3$ is finite for any mass ratio. For three identical bosons, the contact $C_3^{'}$ vanishes, as already shown in Ref.~\cite{castindensity}. In this case, the sub-leading contribution to  the momentum distribution  presents a log-periodic oscillation around zero. For all the other cases where $m_A\neq m_B$, we found that $C_3^{'}$ is finite, which is in agreement with the results obtained numerically  in Ref.~\cite{yamashita2013}.

\begin{center}
\begin{figure}[t] 
{\includegraphics[width=8.5cm]{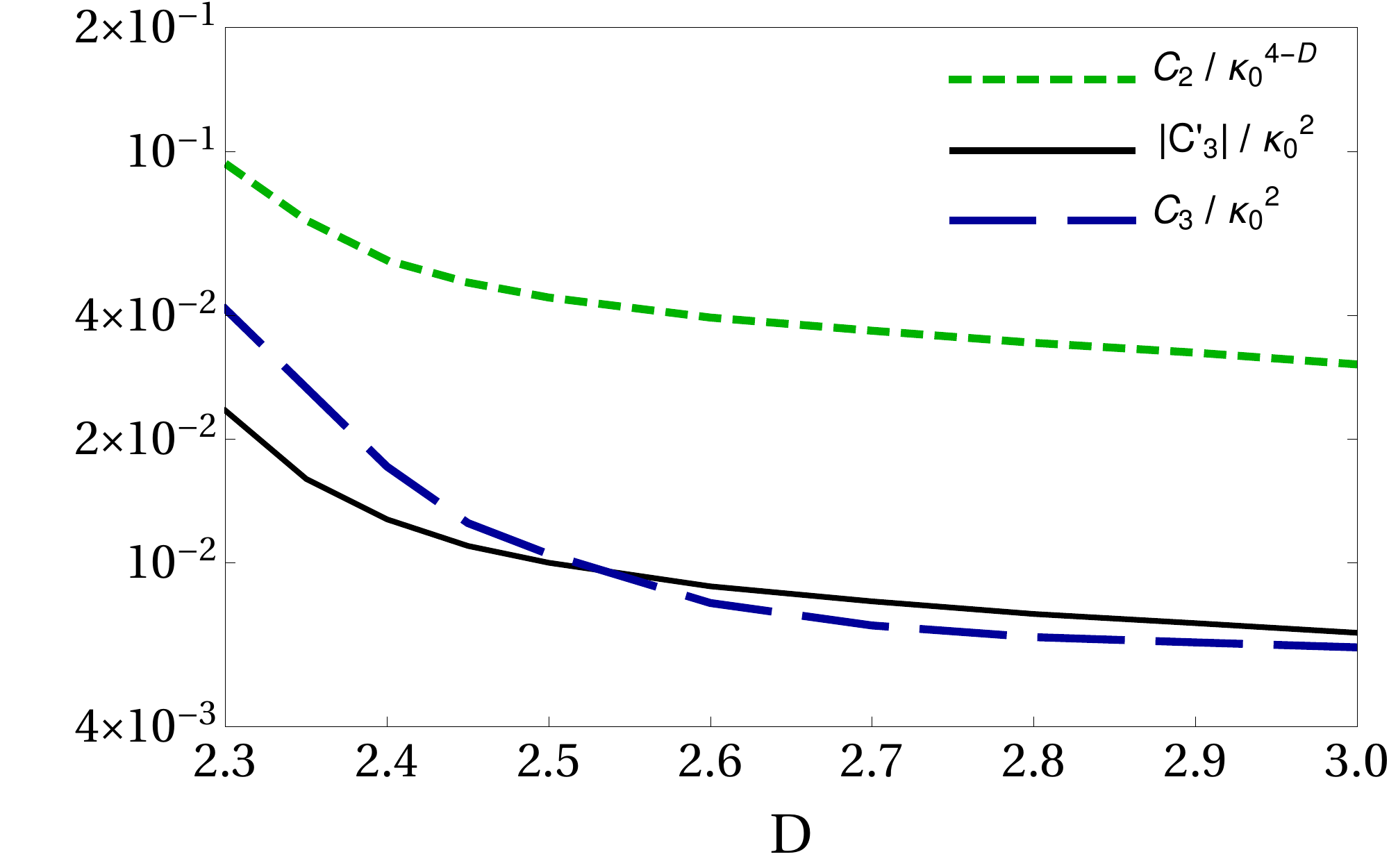}} 
{\includegraphics[width=8cm]{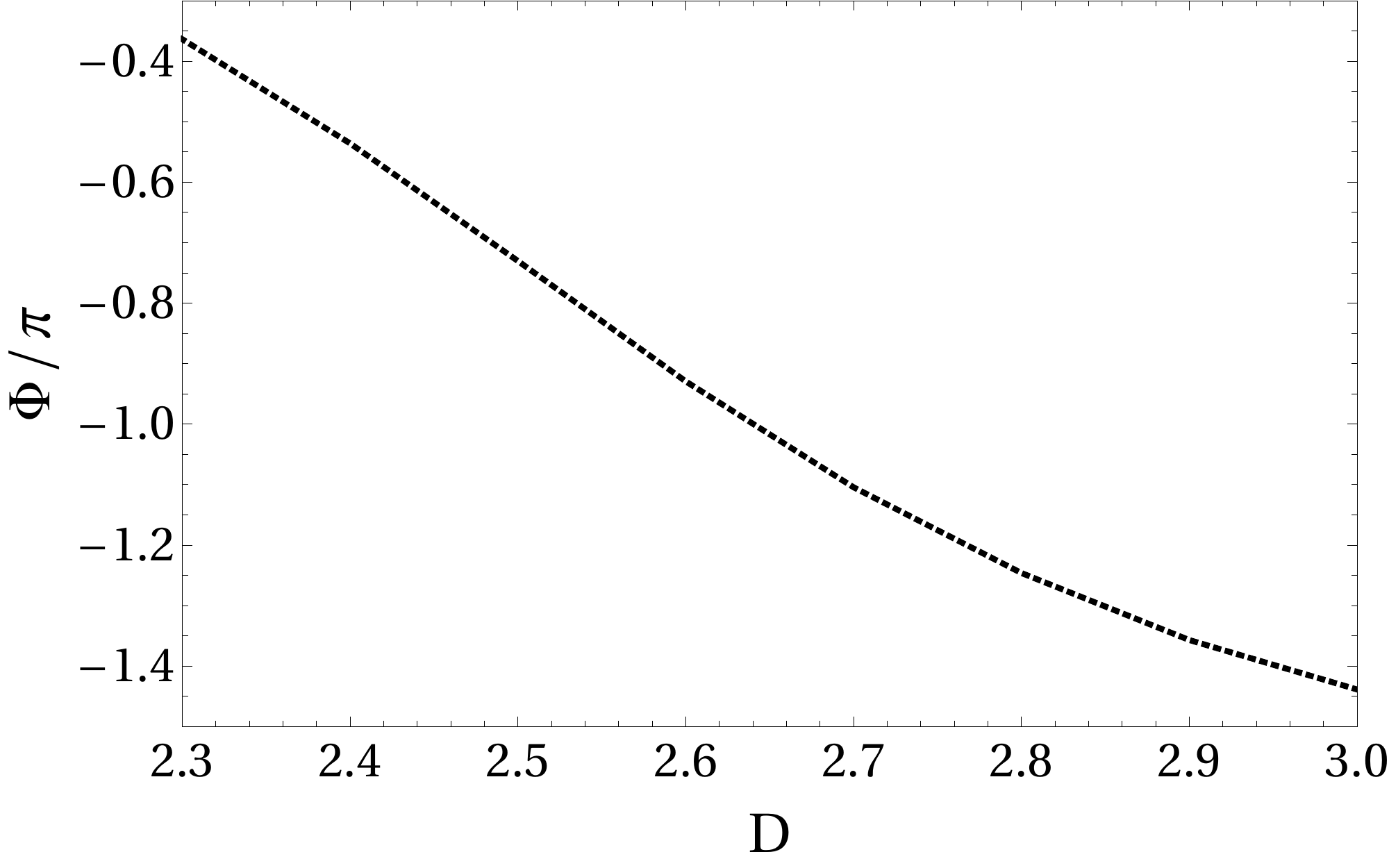}} 
\caption{ Three- and two-body contact parameters  and phase  for  the   $^{6}$Li$-^{133}$Cs$_{2}$ system in noninteger  dimensions from 2.3 to three.  
 Top panel: ${C_3'}/{\kappa_0^2}$ (solid line), ${|C_3|}/{\kappa_0^2}$
 (long-dashed line) and ${|C_2|}/{\kappa_0^{4-D}}$ (short-dashed line). Lower panel: phase $\Phi/\pi$ (dotted line).}
\label{fig5}
\end{figure}
\end{center}

In Fig.~\ref{fig5}, we show the dependence of the contacts and phase with the 
noninteger dimension for the $^{6}$Li$-^{133}$Cs$_{2}$ system from noninteger dimension 2.3 up to  three. In the top panel of the figure, we observe that the contacts decrease by moving from 2.3 to three dimensions, which can be understood  as the system turns to be  more dilute for a fixed binding energy as the dimension increases, and in this particular case,  the $^6$Li is less probable to be found at short distances as one increases the dimension. Therefore, the asymptotic tail for large momentum is depleted by increasing dimension, which is reflected in the lowering of the contacts. To be complete, we present in the lower panel the phase as a function of the dimension. 

\begin{center}
\begin{figure}[t]
{\includegraphics[width=8.4cm]{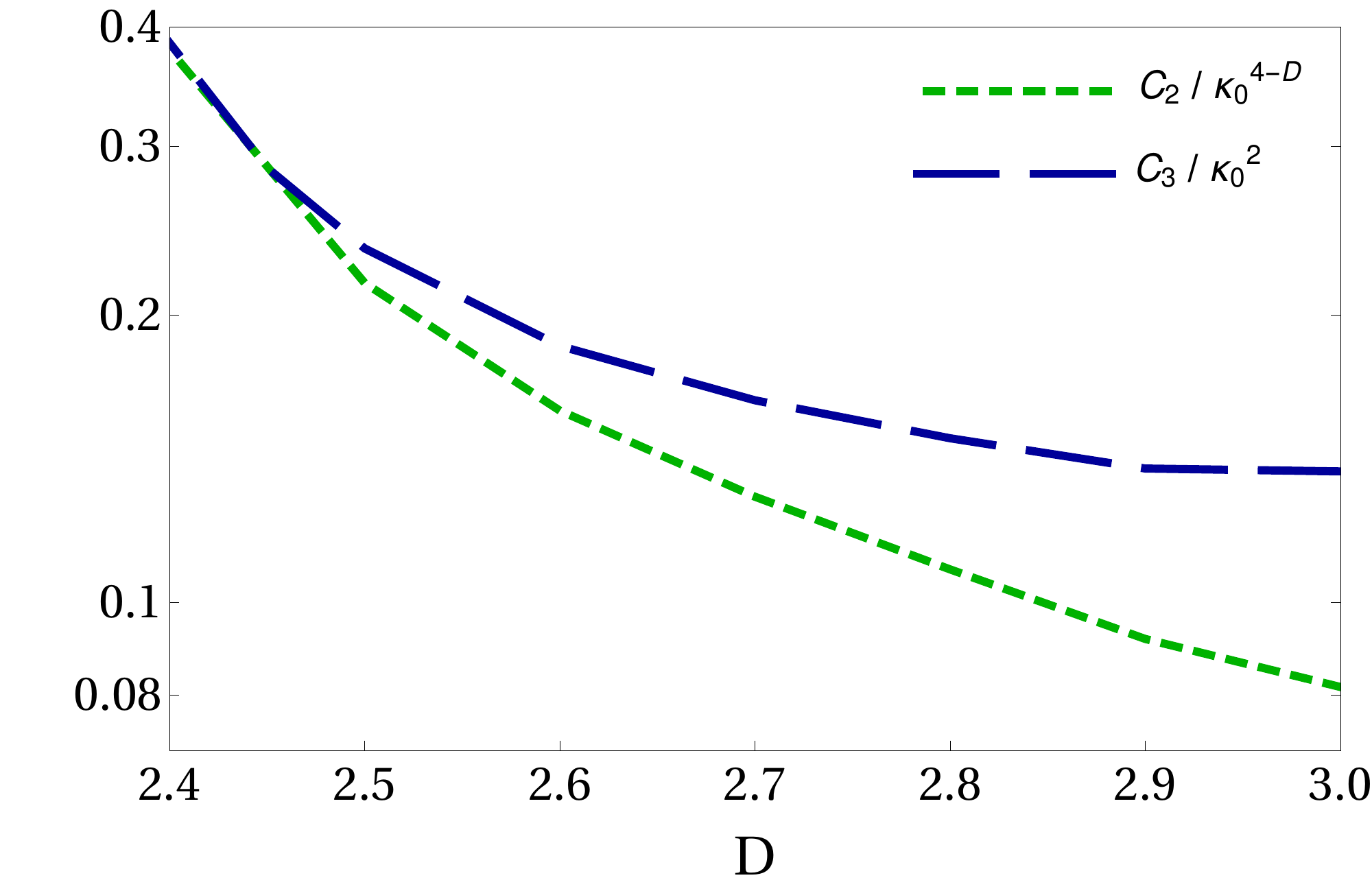}}
\vspace{0.3cm}
{\includegraphics[width=8.cm]{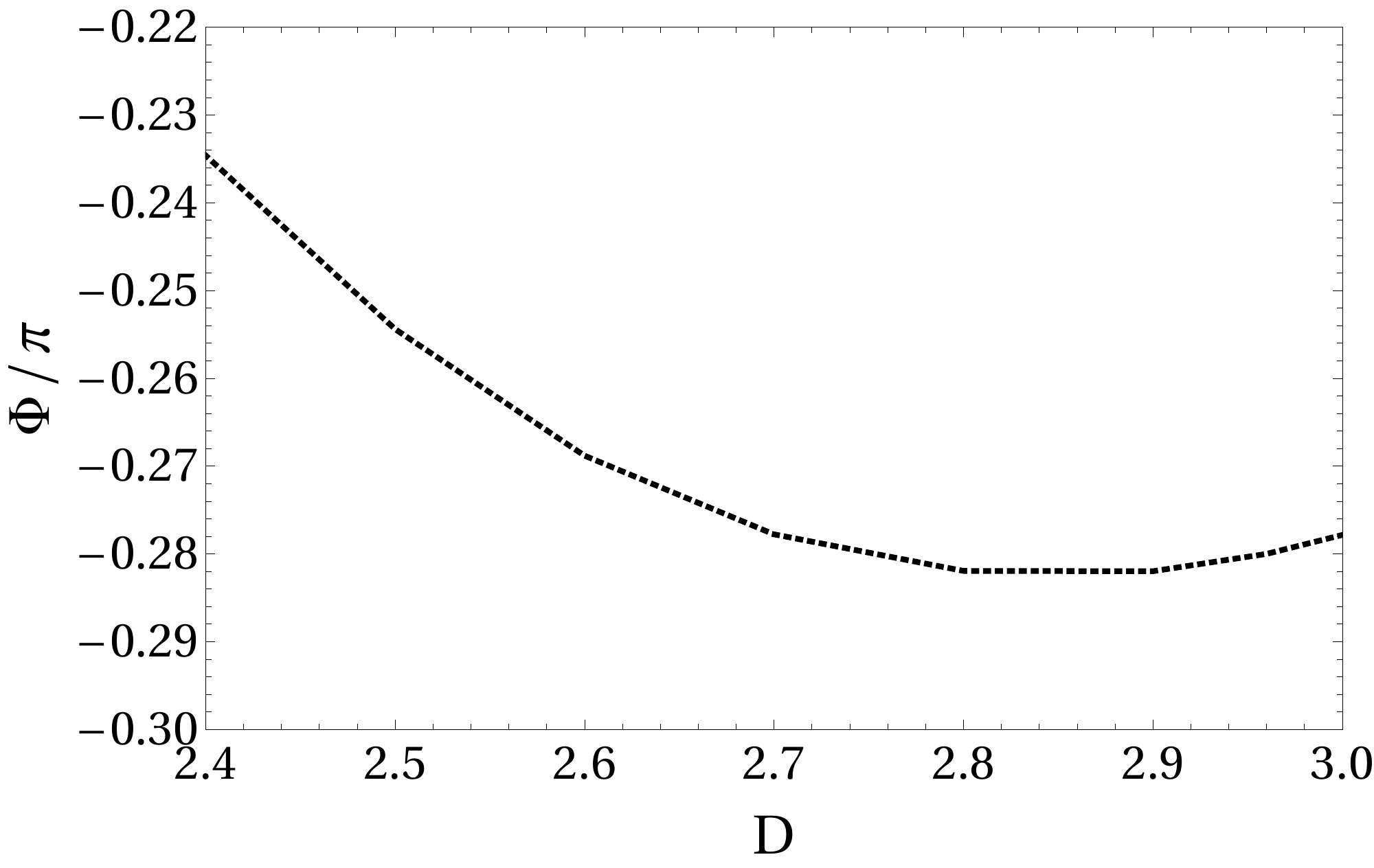}}
\caption{Three- and two-body contact parameters  and phase 
 for  three-identical bosons in noninteger  dimensions.  
 Top panel:  ${C_3}/{\kappa_0^2}$ 
 (long-dashed line) and ${C_2}/{\kappa_0^{4-D}}$ (short-dashed line). Bottom panel:  phase $\Phi/\pi$ (dotted line).}
\label{fig6}
\end{figure}
\end{center}

Finally, in Fig.~\ref{fig6}, we consider the case of  three resonantly interacting identical bosons. We change the noninteger dimension from 3 to $D=2.4$. In this case, the Efimov effect is present until the critical dimension of $D_c=2.3$, which corresponds to a squeezed trap with $b_{ho}/r_{2D}=\sqrt{0.994}$. We observe in the figure, that $C'_3$ is always zero, while towards the critical dimension the two- and three-body contacts increase,  and the phase approaches the value of $-0.737$ for $D=2.4$ or squeezed trap with $b_{ho}/r_{2D}=\sqrt{1.429}$. 

\section{SUMMARY}
 \label{section4}

In this work, we calculated the single particle momentum distribution of an Efimov 
mass-imbalanced state in noninteger dimensions at unitarity. We used the wave 
function of an Efimov state with a finite three-body binding energy, obtained 
previously in Ref.~\cite{betpeiPRA} - in that reference, the three-body energy 
eigenstate was derived in configuration space by considering the Bethe-Peierls 
boundary conditions in the limit of a zero-range interaction and  
infinite two-body scattering length. 

We studied the single particle momentum distribution in terms of the relative 
momentum of particle $B$ with respect to the $AA$ subsystem. For that, the Fourier 
transform of the Efimov state wave function was performed relying on the spectator 
functions obtained analytically by the application of the free resolvent to each 
Faddeev component of the wave function, following the method developed in 
Ref.~\cite{castindensity}. These spectator functions depend only on the relative
momentum of the spectator particle to the center of mass of the interacting pair. 
They have the characteristic log-periodic oscillations at large momentum, which 
depend on the noninteger dimension. Furthermore, due to the finite three-body 
binding energy, the spectator functions are finite for vanishing momentum. Their
analytical form reproduces the known numerical results from the 
literature~\cite{yamashita2013}.

The task of deriving the leading and sub-leading contributions to the high 
momentum tail of the single particle momentum density and the associated 
two- and three-body contact parameters were made possible by using the 
analytic form of the spectator functions in momentum space. Independently 
of the noninteger dimension, the leading non-oscillatory large momentum 
tail scales as $1/q^4$ (see Eqs.~\eqref{eq:n2} and \eqref{c3c3l})
and is normalized by the two-body contact. The sub-leading term is 
proportional to $1/q^{D+2}$; it is composed of the sum of two contributions, 
a log-periodic and a non-oscillatory, each one normalized by the 
corresponding three-body contact as shown in Eq.~\eqref{c3c3l}.

The contact parameters were then computed by decreasing the noninteger dimension 
starting from 3D, where the Efimov discrete scaling drives the physics of the 
three-body system, until close to the critical dimension, when the transition to the continuum scale symmetry takes place. We found that the two- and three-body parameters tend to increase in magnitude close to the critical dimension, independently of the mass imbalance of the three-body system. 

We explored in detail the systems formed by $^6$Li$-^{133}$Cs$_2$ and three-identical bosons. The parameter $C'_3$, normalizing the sub-leading 
non-oscillatory term, is zero for three-identical bosons regardless of the 
noninteger dimension. For the $^6$Li$-^{133}$Cs$_2$ system, we found that the two- 
and three-body contact parameters increase close to the critical dimension  where 
the Efimov effect disappears. Furthermore, in this case, the phase of the log-
periodic term approaches $-1.143$ for $D=2.4$. The expectation of the growth of 
the two- and three-body contact parameters with the decrease of the noninteger 
dimension seems natural, as one can naively infer, in this situation, the 
particles have the chance to stay closer to the system confined in an oblate 
trap. 

In summary, we have explored different aspects of  the momentum density of 
particle $B$ for mass-imbalanced $AAB$ systems in noninteger dimensions, 
which can be a useful probe into the effect of trap deformation in few-body dynamics 
and in the Efimov phenomenon. The hallmark of this transition from three 
to the critical dimension where the Efimov effect vanishes can be seen  
in the asymptotic momentum distribution, where we show that the contact 
parameters grow, and, consequently, their effects in the evolution of the 
many-body properties by decreasing the noninteger dimension.


\section*{ACKNOWLEDGMENTS}
This work was partially supported by Funda\c{c}\~ao de Amparo \`a Pesquisa do 
Estado de S\~ao Paulo (FAPESP) [grant nos. 2017/05660-0 and 2019/07767-1 (T.F.), 
2020/00560-0 (D.S.R.), and  2018/25225-9 (G.K.)], and Conselho Nacional 
de Desenvolvimento Cient\'{i}fico e Tecnol\'{o}gico (CNPq) [grant nos. 
308486/2015-3 (T.F.), 303579/2019-6 (M.T.Y.), and  309262/2019-4 (G.K.)]. 

\appendix

 \section{Sub-leading contributions to $n_1(q_B)$}
\label{appn1}
 
 Eq.~\eqref{n1} can be written in spherical coordinates as
\begin{equation}
 n_{1}(q_B) =  \lvert \chi^{(B)}(q_B) \rvert^{2} \mathcal{S}_{D} \int^\infty_0 \hspace{-.3cm}d p_B \frac{p_{B}^{D-1}}{\left(E_3 + p_{B}^{2}+\frac{q_{B}^{2}}{2\mu_B} \right)^{2}}, 
 \end{equation}
 where $\mathcal{S}_D = 2\pi^{D/2}/\Gamma(D/2)$. Changing variables $p_{B}/q_{B} = p'_B$ and considering $q_B\gg \sqrt{2\mu_BE_3}$, allow us to write
 \begin{eqnarray}
 n_{1}(q_B) &=&  \frac{\lvert \chi^{(B)}(q_B) \rvert^{2}}{q_{B}^{4-D}}\mathcal{S}_{D} \int^\infty_0 d p'_B \frac{p'^{D-1}_B}{\left( p'^{2}_B+ 1/2\mu_B\right)^{2}} \nonumber \\
&=&\frac{\lvert \chi^{(B)}(q_B) \rvert^{2}}{q_{B}^{4-D}}\mathcal{S}_{D} 
\frac{(2-D)\pi}{4} \csc\left( \frac{D\pi}{2} \right)\left(2\mu_B\right)^{2-D/2}\,.\nonumber \\
 \end{eqnarray}
 
For large momentum, we use the asymptotic spectator function, Eq.~\eqref{eq:assymp}, and from simple manipulations we separate the oscillatory term, namely log-periodic one:
\begin{eqnarray}\label{eq:A3}
 n^\text{osc}_{1}(q_B)&=&  \frac{\left| C^{(B)}\right|^{2}}{q_B^{D+2}}\cos\left[2s_0 \log\left( \frac{q_B}{\sqrt{2\mu_B}\kappa_0^{*}}\right)\right] |\mathfrak{F}_{(D,s_0)}|^2 \nonumber \\ 
 &\times&\mathcal{S}_{D}\pi\left(1-\frac D2\right)\left[\operatorname{Re}(\mathcal{G})^2+\operatorname{Im}(\mathcal{G})^2\right] \nonumber \\
 &\times& (2\mu_B)^{1+D/2}\csc\left( \frac{D\pi}{2} \right) ,
 \end{eqnarray}
 and the non-oscillatory part as
  \begin{eqnarray} \label{eq:A4}
 n^\text{nosc}_{1}(q_B)&=&  \frac{\left| C^{(B)}\right|^{2}}{q_B^{D+2}}|\mathfrak{F}_{(D,s_0)}|^2 \mathcal{S}_{D}\pi \left(1-\frac D2\right)(2\mu_B)^{1+D/2}\nonumber \\
 &\times&\left[\operatorname{Re}(\mathcal{G})^2+\operatorname{Im}(\mathcal{G})^2\right] \csc\left( \frac{D\pi}{2} \right)\, ,
 \end{eqnarray}
where $\mathcal G$ is written in Eq.~\eqref{eq:G}.
 
 \section{Sub-leading contributions to $n_2(q_B)$}
\label{appn2}
 
Taking the large momentum limit, where $q_B\gg \sqrt{2\mu_BE_3}$,  and changing variables to $\textbf{q}_A=\textbf{p}_B-\textbf{q}_B/2$, Eq.~\eqref{n2} can be written as:
 \begin{equation}
 n_{2}(q_B) =  2 \int d^{D}q_A \frac{\lvert \chi^{(A)}(q_A) \rvert^{2}}
 { \left( q_A^{2}+ \textbf{q}_A.\textbf{q}_B + q_{B}^{2}/2\mu_B \right)^{2} }\, .
 \end{equation}
 
 In spherical coordinates we have that:
 \begin{eqnarray}
 &&n_{2}(q_B) = \frac{2(2\pi)}{q_{B}^{4}}\prod_{k=1}^{D-3}\int_0^{\pi} d\theta_k\sin^{k}\theta_k \nonumber \\
 &&\times\int^\infty_0 dq_A\   q_A^{D-1}  | \chi^{(A)}(q_A) |^{2}
\int_0^{\pi} d\theta\ \sin^{D-2}\theta \nonumber \\
&&\times\frac{1}{\left[ (q_A/q_B)^{2} + (q_A/q_B)\cos\theta+ (\mathcal{A}+1)/2\mathcal{A} \right]^{2}}\,, \ \ \ \ \ \ 
 \end{eqnarray}
changing variables to $q'_A=q_A/ q_B$, we find that:
 \begin{equation}
 \hspace{-0.3cm}n_{2}(q_B) =  {2\mathcal{S}_D\over q_B^{4-D}} \int^\infty_0 \hspace{-0.1cm}dq^\prime_A \, q_A^{\prime\,D-1} \mid \hspace{-0.1cm}\chi^{(A)}(q_B\, q'_A) \hspace{-0.1cm}\mid^{2} \mathcal{H}(q'_A), \label{eq:B3}
 \end{equation}
 with
 \begin{small}
  \begin{multline}
 \mathcal{H}(y)=\frac{4 \mathcal{A}^2 (D-2)}{\mathcal{A}^2 \left(4 y^4+1\right)+\mathcal{A} \left(4 y^2+2\right)+1} \\
 +\frac{4 \mathcal{A}^2 (3-D) \left(2  y^2\ \mathcal{A}+\mathcal{A}+1\right) \, }{[2 \mathcal{A} (y-1) y+\mathcal{A}+1]^2 [2  y (y+1)\mathcal{A}+\mathcal{A}+1]} \\
 \times H_2F_1\left(1,\frac{D-1}{2},D-1,-\frac{4 \mathcal{A} y}{2 (y-1) y \mathcal{A}+\mathcal{A}+1}\right),
 \label{eq:H2F1}
 \end{multline}
 \end{small}
where $H_2 F_1(a,b,c,z)$ is the hyper-geometrical function. 

In order to separate the oscillatory and non-oscillatory contributions in $n_2(q_B)$~\eqref{eq:B3}, we perform the following manipulation:
 \begin{eqnarray}
n_{2}(q_B) = && \frac{2\mathcal{S}_D }{q_B^{4-D}}\int^\infty_0 dq'_A \, q_A^{\prime D-1} | \chi^{(A)}(q_B\, q'_A) |^{2}\nonumber \\ &\times& \left(\mathcal{H}(q'_A)-\frac{4 \mathcal{A}^2}{(\mathcal{A}+1)^2}\right)+\frac{C_2}{q_B^{4}},
 \label{eq:B5}\end{eqnarray}
where $C_2$ is the two-body contact parameter and was written in Eq.~\eqref{eq:c2}. 

In order to derive the oscillatory part of $n_2(q_B)$ at large momentum, we introduce the asymptotic spectator function, Eq.~\eqref{eq:assymp}, in the first term of Eq.~\eqref{eq:B5}, i.e.,
$n_{2}(q_B) -C_2/q_B^{4}$. The $\cos^2$ from the asymptotic expression of $| \chi^{(A)}(q_B\, q'_A) |^{2}$ is algebraically manipulated
 in the form:
\begin{multline}
\cos^{2}\left[s_0 \log\left({\frac{q'_{A}\ q_B}{\sqrt{2\mu_A}\kappa_0^{*}}}\right)\right]=\frac{1}{2}
\\
+\frac{1}{2}\left\{\cos\left[2s_0 \log\left({\frac{ q_B}{\sqrt{2\mu_A}\kappa_0^{*}}}\right)\right]\cos\left[2s_0 \log(q'_{A})\right]\right. 
\\-\left.\sin\left[2s_0 \log\left(\frac{ q_B}{\sqrt{2\mu_A}\kappa_0^{*}}\right)\right]\sin\left[2s_0 \log({q'_{A})}\right] \right\}.
\end{multline}
we can write the oscillatory
term:
\begin{eqnarray}\label{eq:B7}
&n^\text{osc}_{2}&(q_B) -\frac{C_2}{q_B^{4}}
=\frac{\left| C^{(A)}\right|^{2}}{q_B^{D+2}}\frac{2^{1+D}\mathcal{S}_D}{\mu_A^{1-D}}\left[\operatorname{Re}(\mathcal{G})^2+\operatorname{Im}(\mathcal{G})^2\right]\nonumber \\
&\times&|\mathfrak{F}_{(D,s_0)}|^2\int^\infty_0 dq'_A \ q_A^{\prime 1-D}  \ \left( \mathcal{H}(q'_A)-\frac{4 \mathcal{A}^2}{(\mathcal{A}+1)^2}\right)\nonumber \\
&\times&\left\{\cos\left[2s_0 \log\left({\frac{ q_B}{\sqrt{2\mu_A}\kappa_0^{*}}}\right)\right]\cos\left[2s_0 \log(q'_{A})\right]\right. \nonumber \\
&-&\left.\sin\left[2s_0 \log\left(\frac{ q_B}{\sqrt{2\mu_A}\kappa_0^{*}}\right)\right]\sin\left[2s_0 \log({q'_{A})}\right] \right\}.
\end{eqnarray}
and the non-oscillatory one:
\begin{eqnarray}\label{eq:B8}
&&n^\text{nosc}_{2}(q_B)  -\frac{C_2}{q_B^{4}}
=\frac{\left| C^{(A)}\right|^{2}}{q_B^{D+2}}\frac{2^{1+D}\mathcal{S}_D}{\mu_A^{1-D}}\left[\operatorname{Re}(\mathcal{G})^2+\operatorname{Im}(\mathcal{G})^2\right] \nonumber \\
&&\times|\mathfrak{F}_{(D,s_0)}|^2\int^\infty_0 dq'_A \ q_A^{\prime 1-D}  \ \left( \mathcal{H}(q'_A)-\frac{4 \mathcal{A}^2}{(\mathcal{A}+1)^2}\right)\,.
\end{eqnarray}

 \section{Sub-leading contributions to $n_3(q_B)$}
\label{appn3}
 
Taking $n_3(q_B$) from Eq.~\eqref{n3} with the change of variables $\textbf{p}_B - \textbf{q}_B/2 = \textbf{q}_A$ and considering the large momentum limit, namely $q_B\gg \sqrt{2\mathcal{A}E_3/(\mathcal{A}+1)}$, we can write:
  \begin{equation}
n_{3}(q_B)=  \int d^{D}q_A \frac{2\chi^{(B)}\overset{*}{(}q_B)\ \chi^{(A)}(q_A )  }
 { \left( q_{A}^{2} +\textbf{q}_A.\textbf{q}_B +q_B^{2} \frac{\mathcal{A}+1}{2\mathcal{A}} \right)^{2} } + {\rm c.c.}\,.
 \end{equation}
The spectator functions are real, once again after changing variables $q_A/q_B =q'_A$ and integrating in spherical coordinates, we get that:
\begin{small}
 \begin{equation}
 n_{3}(q_B)= \chi^{(B)}\overset{*}{(}q_B)\frac{4\mathcal{S}_D}{q_B^{4-D}} \int^\infty_0 \hspace{-.3cm}  dq'_A  \ q_A^{\prime D-1}  \chi^{(A)}(q_B q'_A)\mathcal{H}(q'_A)\,,
 \end{equation}
 \end{small}
where $\mathcal{H}(q'_A)$ is  given by Eq.~\eqref{eq:H2F1}. The asymptotic form is found by using the spectator function from Eq.~\eqref{eq:assymp}, leading to:
\begin{eqnarray}
 &&n_{3}(q_B)=\frac{C^{(B)^{\scalebox{0.7}{*}}}C^{(A)} }{q_B^{D+2}} 2^{D+3}\mathcal{S}_D   (\mu_B\mu_A)^{D/2-1/2}\nonumber \\
 &&\times\left[\operatorname{Re}(\mathcal{G})^2+\operatorname{Im}(\mathcal{G})^2\right]\left\{\cos\left[s_0 \log\left(\frac{ q_B}{\sqrt{2\mu_B}\kappa_0^{*}}\right)\right]\right.\nonumber \\
 &&\times\left.\cos\left[s_0 \log\left(\frac{ q_B}{\sqrt{2\mu_A}\kappa_0^{*}}\right)\right]\int^\infty_0 \hspace{-.2cm}dq'_A  \mathcal{H}(q'_A)\cos\left[s_0 \log\left(q'_{A}\right)\right]\right. \nonumber \\
 &&-\left.\sin\left[s_0 \log\left(\frac{ q_B}{\sqrt{2\mu_B}\kappa_0^{*}}\right)\right]\sin\left[s_0 \log\left(\frac{ q_B}{\sqrt{2\mu_A}\kappa_0^{*}}\right)\right]\right.\nonumber \\
 &&\times\left.\int^\infty_0\hspace{-.2cm} dq'_A  \mathcal{H}(q'_A)\sin\left[s_0 \log\left(q'_{A}\right)\right]\right\}\,.
 \end{eqnarray}
 
The algebraically manipulation of the cosines and sines in the equation above leads to  identify the oscillatory 
 term:
 \begin{small}
 \begin{eqnarray}\label{eq:C4}
 &&n^\text{osc}_{3}(q_B)=\frac{C^{(B)^{\scalebox{0.7}{*}}}C^{(A)}  }{q_B^{D+2}} 2^{D+2}\mathcal{S}_D   (\mu_B\mu_A)^{D/2-1/2}  \nonumber \\
 &&\times |\mathfrak{F}_{(D,s_0)}|^2\left[\operatorname{Re}(\mathcal{G})^2+\operatorname{Im}(\mathcal{G})^2\right] \nonumber \\
 &&\times\left\{\cos\left[s_0 \log\left(\frac{ q_B^{2}/\kappa_0^{*2}}{2\sqrt{\mu_A\mu_B}}\right)\right]\int^\infty_0 \hspace{-.2cm}dq'_A  \mathcal{H}(q'_A)\cos\left[s_0 \log\left(q'_{A}\right)\right]\right. \nonumber \\
 &&-\left.\sin\left[s_0 \log\left(\frac{ q_B^{2}/\kappa_0^{*2}}{2\sqrt{\mu_A\mu_B}}\right)\right]
 \int^\infty_0 \hspace{-.2cm}dq'_A  \mathcal{H}(q'_A)\sin\left[s_0 \log\left(q'_{A}\right)\right]\right\}, \nonumber \\
 \end{eqnarray}
 \end{small}
 and the non-oscillatory one as:
 \begin{small}
 \begin{eqnarray}\label{eq:C5}
 &&n^\text{nosc}_{3}(q_B)=\frac{C^{(B)^{\scalebox{0.7}{*}}}C^{(A)}  }{q_B^{D+2}} 2^{D+2}\mathcal{S}_D   (\mu_B\mu_A)^{D/2-1/2} \nonumber \\ 
 &&\times |\mathfrak{F}_{(D,s_0)}|^2\left[\operatorname{Re}(\mathcal{G})^2+\operatorname{Im}(\mathcal{G})^2\right] \nonumber \\
 &&\times\left\{\cos\left[s_0 \log\left(\sqrt{\frac{\mu_B}{\mu_A}}\right)\right] \int^\infty_0 \hspace{-.2cm} dq'_A  \mathcal{H}(q'_A)\cos\left[s_0 \log\left(q'_{A}\right)\right]\right. \nonumber \\
 &&-\left.\sin\left[s_0 \log\left(\sqrt{\frac{\mu_B}{\mu_A}}\right)\right] \int^\infty_0 \hspace{-.2cm} dq'_A  \mathcal{H}(q'_A)\sin\left[s_0 \log\left({q'_{A}}\right)\right]\right\}. \nonumber \\
 \end{eqnarray}
 \end{small}

\section{Sub-leading contributions to $n_4(q_B)$}
\label{appn4}

The  argument of the spectator function in  Eq.~\eqref{n4} is
  \begin{eqnarray}
 \lvert \textbf{p}_B\pm\frac{\textbf{q}_B}{2}\rvert = q_B \sqrt{\frac{p_B^{2}}{q_B^{2}}+\frac{1}{4}\pm \frac{p_B}{q_B}\cos\theta},
 \end{eqnarray}
 then changing variables to $p_B /q_B =  p'_B$ and considering the large momentum limit, one has:
 \begin{eqnarray}\label{eq:D2}
 n_{4}(q_B) &= & \frac{1}{q_B^{4-D}} 4\pi\prod_{k=1}^{D-3}\int_0^{\pi}d\theta_k  \sin^{k}(\theta_k)\nonumber   \\
 &\times&\int^\infty_0 dp'_B  \frac{p_B^{\prime D-1}  }
 { \left[ p_{B}^{\prime 2} + (\mathcal{A}+2)/4\mathcal{A} \right]^{2} } \int_0^{\pi}d\theta \sin^{D-2}\theta \nonumber  \\
 &\times& \chi^{(A)}\overset{*}{(} q_B \, p'_{B -})\chi^{(A)}(q_B \, p'_{B +})\, ,
 \end{eqnarray}
 where $ p'_{B\pm}=\sqrt{p^{\prime 2}_B+\frac14\pm p'_B\cos\theta}$. The product of the spectator functions~\ref{eq:assymp} allow us to write
 \begin{eqnarray}
 &&\chi^{(A)}\overset{*}{(}q_B \ p'_{B -})\chi^{(A)}(q_B \ p'_{B +}) =2 \left[\operatorname{Re}(\mathcal{G})^2+\operatorname{Im}(\mathcal{G})^2\right] \nonumber \\
 &&\times  |\mathfrak{F}_{(D,s_0)}|^2  \left(\frac{  q_B}{\sqrt{2\mu_A}}\sqrt{p'_{B-}\ p'_{B+}}\right)^{2-2D}\nonumber \\
 &&\times \left\{ \cos\left[s_0 \log\left(\frac{  p'_{B+}}{p'_{B-}}\right)\right] +\cos\left[s_0 \log\left(\frac{ q_B^{2}\ p'_{B+}p'_{B-}}{2\mu_A\kappa_0^{*\ 2}}\right)\right]\right\}.\nonumber \\
 \end{eqnarray}
Then, the oscillatory contribution in Eq.~\eqref{eq:D2} is:
   \begin{eqnarray}\label{eq:D4}
 &&n^\text{osc}_{4}(q_B)=\frac{|C^{(A)}|^{2}}{q_B^{D+2}}  \frac{2^{2+D}\pi^{D/2-1/2}}{\Gamma[D/2-1/2]}|\mathfrak{F}_{(D,s_0)}|^2\mu_A^{D-1}\nonumber \\
 &&\times\left[\operatorname{Re}(\mathcal{G})^2+\operatorname{Im}(\mathcal{G})^2\right] \nonumber \\ 
 &&\times  \int^\infty_0 dp'_B \frac{p_B^{\prime D-1}  }
 { \left[ p_{B}^{\prime 2} + (\mathcal{A}+2)/4\mathcal{A} \right]^{2} } \int_0^{\pi}d\theta \sin^{D-2}\theta\nonumber \\ 
  &&\times  \mathcal{W}^{1/2-D/2}  \cos\left[s_0 \log\left( \frac{q_B^2}{2\mu_A \kappa_0^{*2}}\, \mathcal{W}^{1/2}\right)\right]\, ,
 \end{eqnarray}
 where 
 \begin{equation}
  \mathcal{W}=\left(p^{\prime 2}_B+\frac{1}{4}+ p'_B\cos\theta\right)\left(p^{\prime 2}_B+\frac{1}{4}- p'_B\cos\theta\right)\, .\nonumber
 \end{equation}
The non-oscillatory contribution in Eq.~\eqref{eq:D2} can be identified as:
   \begin{eqnarray}\label{eq:D5}
 &&n^\text{nosc}_{4}(q_B)=\frac{|C^{(A)}|^{2}}{q_B^{D+2}}  \frac{2^{2+D}\pi^{D/2-1/2}}{\Gamma[D/2-1/2]}|\mathfrak{F}_{(D,s_0)}|^2\mu_A^{D-1} \nonumber \\
 &&\times\left[\operatorname{Re}(\mathcal{G})^2+\operatorname{Im}(\mathcal{G})^2\right] \nonumber \\ 
  &&\times   \int^\infty_0 dp'_B \frac{p_B^{\prime D-1}  }
 { \left[ p_{B}^{\prime 2} + (\mathcal{A}+2)/4\mathcal{A} \right]^{2} } \int_0^{\pi}d\theta \sin^{D-2}\theta\nonumber \\ 
  &&\times  \mathcal{W} ^{1/2-D/2}   \cos\left[s_0 \log\left(  \sqrt{\frac{p^{\prime 2}_B+\frac14+ p'_B\cos\theta}{p^{\prime 2}_B+\frac14- p'_B\cos\theta}} \right)\right]\, . \ \ \ \ \ \ 
 \end{eqnarray}


\end{document}